\documentclass[letterpaper, 10 pt, journal, twoside]{IEEEtran}

\usepackage{graphicx} % for pdf, bitmapped graphics files
\usepackage{epsfig} % for postscript graphics files
\usepackage{subfigure}
\usepackage{amsmath} % assumes amsmath package installed
\usepackage{amssymb}  % assumes amsmath package installed
\usepackage{epstopdf}
\usepackage{lipsum,amsmath,multicol}
\usepackage{float}
\usepackage{algorithm}
\usepackage{hyperref}
\usepackage{algpseudocode}
\usepackage{wrapfig}
\usepackage{sidecap}

\usepackage{amsthm}
\newtheorem{theorem}{Theorem}
\usepackage{comment}
\usepackage{array}
\usepackage{glossaries}
\usepackage{url}
\usepackage{multirow}
\usepackage{tabularx}
\begin{document}
\title{Reactive Navigation under Non-Parametric Uncertainty through Hilbert Space Embedding of Probabilistic Velocity Obstacles}

\author{P. S. Naga Jyotish*$^1$, Bharath Gopalakrishnan*$^1$, A. V. S. Sai Bhargav Kumar$^1$, Arun Kumar Singh$^2$,\\ K. Madhava Krishna$^1$ and Dinesh Manocha$^3$.
\vspace{-1.1cm}
    \thanks{$^1$ are with RRC, IIIT Hyderabad, India {\tt\small mkrishna@iiit.ac.in}, {\tt\small srisai.poonganam@research.iiit.ac.in}, {\tt\small bharathg91@gmail.com}, {\tt\small bhargavvk18@gmail.com}}
    \thanks{$^2$ is with University of Tartu, Estonia {\tt\small aks1812@gmail.com}}
    
    \thanks{$^3$ is with University of Maryland {\tt\small dm@cs.umd.edu}}
    \thanks{*Equal contribution from first two authors.}
    \thanks{This research in part was supported by the IT Academy of Estonia grant (SLTTI19605T) to Arun Kumar Singh}

    % \thanks{Digital Object Identifier (DOI): see top of this page.}
}
%Use only for final RAL version.

% \markboth{IEEE Robotics and Automation Letters. Preprint Version. Accepted January, 2020}
% {Jyotish \MakeLowercase{\textit{et al.}}: Reactive navigation using RKHS} 

\onecolumn
\thispagestyle{empty}

© 2020 IEEE.  Personal use of this material is permitted.  Permission from IEEE must be obtained for all other uses, in any current or future media, including reprinting/republishing this material for advertising or promotional purposes, creating new collective works, for resale or redistribution to servers or lists, or reuse of any copyrighted component of this work in other works.

\twocolumn

\maketitle
% \vspace{-4cm}
\begin{abstract}
The probabilistic velocity obstacle (PVO) extends the concept of velocity obstacle (VO) to work in uncertain dynamic environments. In this paper, we show how a robust model predictive control (MPC) with PVO constraints under non-parametric uncertainty can be made computationally tractable. At the core of our formulation is a novel yet simple interpretation of our robust MPC as a problem of matching the distribution of PVO with a certain desired distribution. To this end, we propose two methods. Our first baseline method is based on approximating the distribution of PVO with a Gaussian Mixture Model (GMM) and subsequently performing distribution matching using  Kullback Leibler (KL) divergence metric. Our second formulation is based on the possibility of representing arbitrary distributions as functions in Reproducing Kernel Hilbert Space (RKHS). We use this foundation to interpret our robust MPC as a problem of minimizing the distance between the desired distribution and the distribution of the PVO in the RKHS. Both the RKHS and GMM based formulation can work with any uncertainty distribution and thus allowing us to relax the prevalent Gaussian assumption in the existing works. We validate our formulation by taking an example of 2D navigation of quadrotors with a realistic noise model for perception and ego-motion uncertainty. In particular, we present a systematic comparison between the GMM and the RKHS approach and show that while both approaches can produce safe trajectories, the former is highly conservative and leads to poor tracking and control costs. Furthermore, RKHS based approach gives better computational times that are up to one order of magnitude lesser than the computation time of the GMM based approach.
\end{abstract}

% \begin{IEEEkeywords}
% IEEE, IEEEtran, journal, \LaTeX, paper, template.
% \end{IEEEkeywords}
% \begin{IEEEkeywords}
% List of keywords (from the RA Letters keyword list)
% \end{IEEEkeywords}

\IEEEpeerreviewmaketitle

\vspace{-0.45cm}
\vspace{-0.2cm}
\section{Introduction}
\IEEEPARstart{R}{eactive} navigation in real-world environments requires robots to consider perception and ego-motion uncertainty explicitly while computing control inputs to ensure robust collision avoidance. Incorporating only the mean information of the uncertainty often turns to be inadequate. In contrast, bounding volume approaches like \cite{bounding_volume1}, \cite{bounding_volume2} are simple and induce robustness but tend to be overly conservative  \cite{prvo}. Under uncertainty, we would typically like to compute low cost control inputs (based on some metric) and at the same time, ensure some upper bound on the risk of collision. In this paper, we formulate these requirements in a robust Model Predictive Control (MPC) framework, wherein the robustness stems from the constraints imposed by Probabilistic Velocity Obstacle (PVO) \cite{pvo}. Essentially, PVOs are chance constraints defined over the deterministic VO \cite{vo}. Thus, a robust MPC formulation is, in fact, an instance of chance-constrained optimization (CCO). Recently CCO has been used as a general template for developing numerous navigation algorithms for a wide class of robots, ranging from aerial vehicles \cite{aerial_chance1} to autonomous cars \cite{chance_ad1}. % \cite{chance_ad2}, \cite{chance_ad3},
Although CCO provides a rigorous template for decision making under uncertainty, they are, in general, computationally intractable. In fact, under non-Gaussian or non-parametric uncertainty, it is difficult to even compute an analytical expression for the chance constraints (such as PVO in our case). 

\vspace{-0.4cm}
\subsection{Contributions and Overview of the Proposed Approach}
Our primary motivation in this paper is to reformulate CCO as a more tractable problem without making any assumption on the parametric form of the underlying uncertainty and/or resorting to any linearization of the constraints. To this end, we show that our robust MPC or CCO, in general, can be interpreted as a problem of distribution matching \footnote{Although conceptually simple, to the best of our knowledge, there are no other works based on this interpretation.}. To be precise, we construct a certain desired distribution and ensure that the distribution of the PVO matches the desired distribution by choosing low-cost control inputs. We present two methods to accomplish this. Our first baseline method builds upon existing works like \cite{manocha_gmm}, \cite{nathan_michael_gmm}, and is based on approximating the desired distribution and PVO as a GMM and subsequently sampling control inputs to minimize the KL divergence between the two. Although this approach produces trajectories with a high probability of collision avoidance, it is computationally expensive and leads to conservative results with high tracking error and control costs.

Our second method is inspired by the result that local similarity of two given distributions can be bounded through their higher order moments \cite{moment_bound_lindsay}. To accomplish moment matching in a tractable manner, we embed our desired distribution and PVO in Reproducing Kernel Hilbert Space (RKHS)\cite{scholkopf}, \cite{scholkopf2} wherein evaluating the similarity of higher-order moments of two given distribution can be performed through Kernel trick. The RKHS based approach provides following advantages

\begin{itemize}

\item It takes the form of a standard non-linear regression problem that can be solved easily through any gradient-based optimizer. Using the kernel trick, the evaluation of the cost or gradients reduces to simple matrix multiplication that can be accelerated using GPUs for large matrix sizes.

\item The RKHS based approach provides an additional tuning parameter for trading off cost and robustness. The parameter has a clear physical interpretation and thus easier to choose.

\item The RKHS based approach leads to substantially lower tracking cost and control cost than GMM, which in turn translates to lower time of traversal to the goal. The former also requires up to 8 times less computation time since it by-passes the need for performing expensive GMM fits. 

\end{itemize}

\vspace{-0.4cm}
\section{Problem Formulation} \label{problem Formulation}
Our Approach assumes a robot operating in 2D space, although extending the method to 3D is straightforward. Some of the essential symbols and notations are summarized in Table \ref{symbols}. We also define some notations in the first place of their use. We use regular faced lower case letters to represent scalars, while boldface variants would represent vectors. Upper faced letters represent the matrices.

\vspace{-0.35cm}
\subsection{Ego-motion uncertainty} We model  the state $\boldsymbol{\xi}_t=(x_t, \dot{x}_t, y_t, \dot{y}_t)$ at any time $t$ as a random variable with an unknown probability distribution. The state evolves according to the following stochastic linear motion model:

\small
\vspace{-0.3cm}
\begin{equation}
\boldsymbol{\xi}_{t} = \textbf{A}\boldsymbol{\xi}_{t-1}+\textbf{B}(\textbf{u}_{t-1}+\boldsymbol{\delta}_{t-1}),
\label{ego_uncert}
\vspace{-0.1cm}
\end{equation}
\normalsize

\noindent where, $\boldsymbol{\delta}_{t-1}$ is a random variable with unknown probability distribution that acts as a perturbation to the control input $\textbf{u}_{t-1}$. We assume that we do not have access to the probability distribution of $\boldsymbol{\delta}_{t-1}$ but we have a black-box process model which can generate different instances of $\boldsymbol{\delta}_{t-1}$. Now, given 
$n$ samples of $\boldsymbol{\xi}_t$ at $t=0$, we can adopt a particle filter like approach and motion model (\ref{ego_uncert}) to propagate uncertainty over time. Note that since we don't know the probability distribution of $\boldsymbol{\xi}_t$ and $\boldsymbol{\delta}_t$, a Kalman Filter like approach is untenable even for a linear system like (\ref{ego_uncert}).

\vspace{-0.35cm}
\subsection{Perception uncertainty} We assume that we have access to a predicted trajectory for each dynamic obstacle in the environment. Consequently, at any time instant $t$, we have access to the state of the $j^{th}$ obstacle $_{j}^{o}\boldsymbol{\xi}_t = ( {_{j}^{o}}x_t, {_{j}^{o}}\dot{x}_t, {_{j}^{o}}y_t, {_{j}^{o}}\dot{y}_t)$. We model perception uncertainty by treating $_{j}^{o}\boldsymbol{\xi}_t$ as random variables with unknown probability distributions. As with ego-motion uncertainty, we assume that we can have access to $n$ samples of $_{j}^{o}\boldsymbol{\xi}_t$. For example, this can come from a separate particle filter which tracks obstacle state over time and is initialized with $n$ samples of $_{j}^{o}\boldsymbol{\xi}_t$ at $t=0$.

%Again, we use a particle filter like approach to evolve $_{j}^{o}\boldsymbol{\xi}_t$ over time.

\vspace{-0.35cm}
\subsection{Velocity Obstacle (VO)} If both the robot and the obstacles are modeled as circular disks, VO is defined by the following set of equations:

\vspace{-0.2cm}
\small
\begin{equation}
f_{t}^j(.)\leq 0:\frac{(\textbf{r}_{j}^T\textbf{v}_{j})^2}{\Vert \textbf{v}_{j}\Vert^2}-\Vert\textbf{r}_{j}\Vert^2+({_{j}^{o}}R+R)^2\leq 0., \forall j, 
\label{vo}
\end{equation}
\normalsize
 
% \vspace{-0.2cm} 
 
\small 
\begin{equation*}
\textbf{r}_{j}= \begin{bmatrix}
    x_t-{_{j}^{o}}x_t  \\
   y_t-{_{j}^{o}}y_t \\
\end{bmatrix} , \textbf{v}_{i} =   \begin{bmatrix}
    \dot{x}_t-{_{j}^{o}}\dot{x}_t  \\
   \dot{y}_t-{_{j}^{o}}\dot{y}_t \\
\end{bmatrix},
\end{equation*}
\normalsize

\noindent where, $R$, ${_{j}^{o}}R$ are the radii of the circular disks representing \newpage the shapes of the robot and obstacle respectively.

\begin{table}[!t]
\centering
\caption{Important Symbols}
\scriptsize
% \vspace{-0.3cm}
\begin{tabular}{|p{1.5cm}|p{5cm}|p{5cm}|p{5cm}|}\hline\label{symbols}
$\boldsymbol{\xi}_{t-1}$ & Robot state at time $t-1$\\ \hline   
${_j^o}\boldsymbol{\xi}_{t}$ & $j^{th}$ Obstacle state at time $t$\\ \hline  
$\boldsymbol{\delta}_{t-1}$ & Control perturbation at time $t-1$\\ \hline  
$\textbf{w}_{t-1}$ & $(\boldsymbol{\xi}_{t-1}, \boldsymbol{\delta}_{t-1} )$\\ \hline  
$\textbf{u}_{t-1}$ & Control input at time $t-1$\\ \hline  
$f_t^j(.)$ & VO constraint under deterministic conditions   \\ \hline
$P_{f_t^j}(\textbf{u}_{t-1})$    & Distribution of $f_t^j(.)$ under uncertainty  \\ \hline
$\eta$ & Chance constraint probability  \\ \hline
$P_{f_t^j}^{des}$     & Desired distribution\\ \hline
$k(.,.)$ & Kernel function\\ \hline
$\mu_{P_{f_t^j}}$    & RKHS embedding of  $P_{f_t^j}$\\ \hline
$\mu_{P_{f_t^j}^{des}}$    & RKHS embedding  of the distribution $P_{f_t^j}^{des}$\\ \hline
\end{tabular}
\normalsize
\vspace{-0.45cm}
\end{table}

\vspace{-0.35cm}
\subsection{Probabilistic Velocity Obstacle (PVO)} 
If the state of the robot evolves according to (\ref{ego_uncert}), then, given the current robot state $\boldsymbol{\xi}_{t-1}$, the predicted next instant obstacle state  ${_{j}^{o}}\boldsymbol{\xi}_{t}$, and the perturbation $\boldsymbol{\delta}_{t-1}$, VO (\ref{vo}) can be represented as an explicit function of the control input.

\small
\vspace{-0.4cm}
\begin{eqnarray}\nonumber
f_{t}^j(.) = h_1(\textbf{w}_{t-1}, {_{j}^{o}}\boldsymbol{\xi}_{t})(u_{t-1}^x)^2+h_2(\textbf{w}_{t-1}, {_{j}^{o}}\boldsymbol{\xi}_{t})u_{t-1}^xu_{t-1}^y\\\nonumber+h_3(\textbf{w}_{t-1}, {_{j}^{o}}\boldsymbol{\xi}_{t})(u_{t-1}^y)^2+h_4(\textbf{w}_{t-1}, {_{j}^{o}}\boldsymbol{\xi}_{t})u_{t-1}^x\\+h_5(\textbf{w}_{t-1}, {_{j}^{o}}\boldsymbol{\xi}_{t})u_{t-1}^y+h_6(\textbf{w}_{t-1}, {_{j}^{o}}\boldsymbol{\xi}_{t})\label{compact_vo},
\end{eqnarray}
\normalsize

\noindent where\footnote{The concatenation of $\boldsymbol{\xi}_{t-1}$ and $\boldsymbol{\delta}_{t-1}$ to obtain $\textbf{w}_{t-1}$ is motivated by how ego-motion uncertainty is evolved using a particle filter like approach. At time $t-1$, we independently sample $n$ samples of $\boldsymbol{\xi}_{t-1}$ and $\boldsymbol{\delta}_{t-1}$ to compute $n$ samples of $\boldsymbol{\xi}_t$ }, $\textbf{w}_{t-1} = (\boldsymbol{\xi}_{t-1}, \boldsymbol{\delta}_{t-t})$,  $u_{t-1}^x, u_{t-1}^y $ are the components of $\textbf{u}_{t-1}$. The functions $h_i(.)$ and $f_t^j(.)$ depend on random variables $\textbf{w}_{t-1}, {_{j}^{o}}\boldsymbol{\xi}_t$ and, consequently, are random variables  with some unknown probability distributions themselves.
To model collision-free velocities under uncertainty, we need to formulate a probabilistic variant of VO. As mentioned earlier, PVOs are essentially chance constraints defined over VO, and thus we have $PVO: P(f_t^j(.)\leq 0)\geq \eta$ where $P(.)$ represents probability.  Essentially, PVO ensures that the VO constraints are satisfied with some lower bound probability, $\eta$. Note that our definition of PVO stems from \cite{pvo} but is written in a slightly different form to highlight it as a chance constraint explicitly. For future use, we use the  notation $P_{f_t^j}(\textbf{u}_{t-1})$ to denote the distribution of $f_t^j(.)$, parameterized in terms of control input $\textbf{u}_{t-1}$ for  given random variables $\textbf{w}_{t-1}, {_{j}^{o}}\boldsymbol{\xi}_t$.

\vspace{-0.4cm}
\subsection{Robust MPC} We formulate the problem of reactive navigation in terms of the following robust MPC. 

\vspace{-0.4cm}
\small
\begin{subequations}
\begin{align}
\arg\min_{\textbf{u}_{t-1}} J(\textbf{u}_{t-1}) = \Vert\boldsymbol{\overline{\xi}}_{t} -   \boldsymbol{\xi}_t^d\Vert_2^2 +\Vert \textbf{u}_{t-1}\Vert_2^2 \label{cost_mpc}\\
\vspace{-0.15cm}
P(f_t^j(\textbf{w}_{t-1},{_{j}^{o}}\boldsymbol{\xi}_{t},  \textbf{u}_{t-1})\leq 0)\geq \eta\label{chance_mpc}\\
\textbf{u}_{t-1} \in \mathcal{C} \label{feasible_mpc}
\end{align}
\vspace{-0.6cm}
\end{subequations}
\normalsize

\noindent where $\overline{\boldsymbol{{\xi}}}_t$ represents the mean of $\boldsymbol{\xi}_t$ and is used to keep the cost function deterministic\footnote{A more rigorous approach would be to define the cost function using the expectation operator. However, the formulation based on the mean state is simpler and acts as a lower bound for the cost based on the expectation operator. \cite{boyd_chance}}. The first term in the cost function makes the mean state track some desired state $\boldsymbol{\xi}_t^d$, which in turn could be defined based on some desired trajectory for the robot. The second term penalizes the usage of control input with a high magnitude. The inequality (\ref{chance_mpc}) ensures the probabilistic safety through the formulation of PVO. $\mathcal{C}$ in (\ref{feasible_mpc}) models the feasible set of the control commands based on velocity and acceleration bounds and is assumed to be convex.

Optimizations of the form (\ref{cost_mpc})-(\ref{feasible_mpc}) are known as chance-constrained optimizations (CCO). Its computational complexity stems from the presence of the chance constraints (\ref{chance_mpc}). As shown in (\ref{compact_vo}), $f_t^j(.)$ are highly non-linear functions of both control inputs $\textbf{u}_{t-1}$ and the random variables $\textbf{w}_{t-1}, {_{j}^{o}}\boldsymbol{\xi}_{t} $. In such a case, it is difficult to even compute an analytical expression for $P(f_t^j(.)\leq 0)$.  In the next section, we present our main theoretical results, which allow reformulating (\ref{cost_mpc})-(\ref{feasible_mpc}) into a tractable, non-linear optimization problem while retaining low sample complexity. 

%As mentioned earlier, existing works replace constraints like (\ref{chance_mpc}) with some tractable surrogate. However, without the knowledge of the probability distribution  of $\textbf{w}_{t-1}, {_{j}^{o}}\boldsymbol{\xi}_{t-1} $ such surrogates needs to be approximated through sampling, in which case sample complexity becomes important.

\begin{figure}[!t]
\centering
\includegraphics[width=0.8\linewidth]{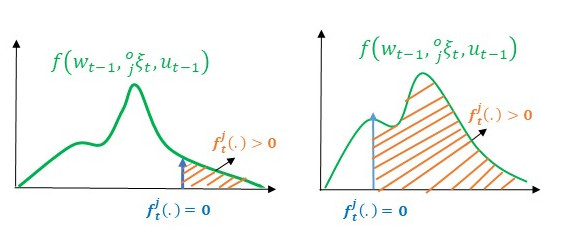}
% \vspace{-0.3cm}
\caption{Intuitive understanding of robust MPC. The shape of the distribution of $f_t^j(.)$ can be manipulated by the control inputs $\textbf{u}_{t-1}$. An appropriate shape is one where most of the mass lies to the left of $f_t^j(.) = 0$ (left figure)}
\vspace{-0.3cm}
\label{intro_dist_fig1}
\vspace{-0.3cm}
\end{figure}

\vspace{-0.35cm}
\section{Main Results}
%
%\subsection{Overview of RKHS based Approach}
%\noindent Fig. \ref{intro_dist_fig} gives an overview of our RKHS based algorithm. We assume access to an environment that provides robot and obstacle states along with control perturbations. This information leads us to random variables $\textbf{w}_{t-1}, {_j^o\boldsymbol{\xi}_t}$ defined in the previous section. Our algorithm begins by computing a subset of these samples denoted by $\widetilde{\textbf{w}}_{t-1}, {_j^o\widetilde{\boldsymbol{\xi}}_t}$. We use these subset samples to estimate a certain desired distribution. Next, we compute the RKHS embedding of the desired distribution and $P_{f_t^j}(\textbf{u}_{t-1})$ through polynomial kernels. Subsequently, we present the minimization of the distance between the two RKHS embedding as a tractable substitute for our robust MPC (\ref{cost_mpc})-(\ref{feasible_mpc}). Our algorithm also uses concepts from reduced sets to improve sample complexity of RKHS embedding. 

\subsection{Intuitive interpretation of robust MPC}
\noindent At an intuitive level, optimization (\ref{cost_mpc})-(\ref{feasible_mpc}) can be interpreted as a problem of ensuring that $P_{f_t^j}(\textbf{u}_{t-1})$ has such a shape that a specific portion of its mass  lies to the left of $f_t^j(.)=0$ (refer to Fig. \ref{intro_dist_fig1}). The chance constraint probability $\eta$ has a direct correlation with the amount of mass lying to the left of $f_t^j(.)=0$. A larger mass amounts to a higher $\eta$. Given random variables $\textbf{w}_{t-1},{_{j}^{o}}\boldsymbol{\xi}_{t}$, the distribution $P_{f_t^j}(\textbf{u}_{t-1})$ is parametrized by  $\textbf{u}_{t-1}$ and therefore can be used to manipulate the shape of $P_{f_t^j}(\textbf{u}_{t-1})$. However, each choice of $\textbf{u}_{t-1}$ incurs a cost $J(\textbf{u}_{t-1})$ in terms of deviation from the desired trajectory and magnitude of the control inputs.
\vspace{-0.45cm}
\subsection{Desired Distribution}\label{sec:desired-distribution}
\noindent Our goal is to ensure that $P_{f_t^j}(\textbf{u}_{t-1})$ achieves an appropriate shape. To this end, our desired distribution acts as a benchmark for $P_{f_t^j}(\textbf{u}_{t-1})$; in other words, a distribution that it should resemble as closely as possible  for an appropriately chosen $\textbf{u}_{t-1}$. We formalize the notion of desired distribution with the help of the following definitions:

\newtheorem{definition}{Definition}
\small
\begin{definition}\label{def_nominal_sol}
The nominal control input $\textbf{u}_{t-1}^{nom}$ refers to any solution of the optimization (\ref{cost_mpc})-(\ref{feasible_mpc}) associated with a cost $J(\textbf{u}_{t-1}^{nom})$ in the vicinity of that obtained in the deterministic setting. 
\end{definition}

\begin{definition}\label{def_des_dist}
Let $\widetilde{\textbf{w}}_{t-1}$, $\widetilde{{_{j}^{o}}\boldsymbol{\xi}}_{t}$ be random variables that represent the same entity as $\textbf{w}_{t-1}$ and ${_{j}^{o}}\boldsymbol{\xi}_t$ but belonging to a different distributions $P_{\textbf{w}_{t-1}}^{des}$, $P_{{_{j}^{o}}\boldsymbol{\xi}_{t-1}}^{des}$, respectively. Further, when   $\widetilde{\textbf{w}}_{t-1} \sim P_{\textbf{w}_{t-1}}^{des}$, $
\widetilde{{_{j}^{o}}\boldsymbol{\xi}}_{t},\sim P_{{_{j}^{o}}\boldsymbol{\xi}_{t}}^{des} $, then  $f_t^j(\widetilde{\textbf{w}}_{t-1},\widetilde{{_{j}^{o}}\boldsymbol{\xi}}_{t}, \textbf{u}_{t-1}) \sim P^{des}_{f_t^j}$. In such a case, $P^{des}_{f_t^j}$ is called the desired distribution if the following holds for some given nominal solution $\textbf{u}_{t-1}^{nom}$.

% \vspace{-0.5cm}
\small
\begin{equation}
P(f_t^j(\widetilde{\textbf{w}}_{t-1},\widetilde{{_{j}^{o}}\boldsymbol{\xi}}_{t}, \textbf{u}_{t-1}^{nom})\leq 0)\approx 1.0, \widetilde{\textbf{w}}_{t-1} \sim P_{\textbf{w}_{t-1}}^{des}, 
\widetilde{{_{j}^{o}}\boldsymbol{\xi}}_{t},\sim P_{{_{j}^{o}}\boldsymbol{\xi}_{t}}^{des} 
\label{desired_chance}
\end{equation}
\normalsize
\vspace{-0.5cm}
\end{definition}
\normalsize

%\begin{figure}[!t]
%\centering
%\includegraphics[width= 8.35cm] {figs/pipeline_10.png}
%\caption{Overview of our RKHS based algorithm}
%\vspace{-0.5cm}
%\label{intro_dist_fig}
%\end{figure}

Essentially, equation (\ref{desired_chance}) implies that if the uncertainty  belongs to the distribution, $P_{\textbf{w}_{t-1}}^{des}$, $P_{{_{j}^{o}}\boldsymbol{\xi}_{t-1}}^{des}$, then the shape and consequently, the entire mass of the distribution of $f_t^j(\widetilde{\textbf{w}}_{t-1},\widetilde{{_{j}^{o}}\boldsymbol{\xi}}_{t}, \textbf{u}_{t-1})$ can be manipulated to lie almost completely to the left of $f_t^j(.)=0$ by choosing $\textbf{u}_{t-1}=\textbf{u}_{t-1}^{nom}$. This setting represents an ideal case because we have constructed uncertainties so appropriately that we can manipulate the distribution of the chance constraints while incurring a nominal cost. Moving forward, we assume that we have access to the samples $\widetilde{\textbf{w}}_{t-1}$, $\widetilde{{_{j}^{o}}\boldsymbol{\xi}}_{t}$  from the distribution $P_{\textbf{w}_{t-1}}^{des}$, $P_{{_{j}^{o}}\boldsymbol{\xi}_{t-1}}^{des}$ respectively and using it we can construct a desired distribution $P^{des}_{f_t^j}$. We lay out the exact process of achieving this in Appendix.

\vspace{-0.33cm}
\subsection{Reformulating (\ref{cost_mpc})-(\ref{feasible_mpc})} 
\noindent From Def.\ref{def_des_dist} it can be deduced that as we make the distribution of chance constraints (PVO) $P_{f_t^j}(\textbf{u}_{t-1})$ more and more similar to $P^{des}_{f_t^j}$, we ensure that more and more mass of $P_{f_t^j}(\textbf{u}_{t-1})$ gets shifted to the left of $f_t^j(.)=0$. As a result, we satisfy the chance constraints (\ref{chance_mpc}) with higher $\eta$. Building on these insights, we present the following reformulation of (\ref{cost_mpc})-(\ref{feasible_mpc}).

\small
\begin{subequations}
\vspace{-0.33cm}
\begin{align}
\arg\min J(\textbf{u}_{t-1})+\rho\mathcal{L}_{dist}(P_{f_t^j}(\textbf{u}_{t-1}),  P^{des}_{f_t^j})\label{cost_reform_gen}\\
\textbf{u}_{t-1}\in \mathcal{C} \label{feas_reform_gen},
\end{align}
\vspace{-0.6cm}
\end{subequations} 
\normalsize

\noindent where, $\mathcal{L}_{dist}$ is a cost which measures some notion of similarity between the $P_{f_t^j}(\textbf{u}_{t-1})$ and  $P^{des}_{f_t^j}$. It should be noted that the chance constraint probability $\eta$ is not explicitly present in the above optimization. Rather, it is accommodated implicitly through the residual of $\mathcal{L}_{dist}$. Lower the residual, higher the $\eta$. The residual in turn can be controlled through the weight $\rho$. 

We solve (\ref{cost_reform_gen})-(\ref{feas_reform_gen}) through a control sampling approach. We discretize the feasible set $\mathcal{C}$ and evaluate the $J(\textbf{u}_{t-1})+\rho\mathcal{L}_{dist}$ for each control sample from the feasible set. Subsequently, we choose the control input which results in minimum value of the cost (\ref{cost_reform_gen}). The primary complexity of our control sampling approach depends on the form of $\mathcal{L}_{dist}$ or more precisely on the computational ease with which we can evaluate $\mathcal{L}_{dist}(P_{f_t^j}(\textbf{u}_{t-1}),  P^{des}_{f_t^j})$ for a given $\textbf{u}_{t-1}$. In the next couple of sections, we present two choices for it based on KL divergence and RKHS embedding.

%\noindent Optimization (\ref{cost_scenario})-(\ref{feasible_scenario}) is precisely the so-called scenario approach for CCO. Herein, the trade-off between robustness and optimal cost can be achieved by controlling the sample sizes $n_r$ and $n_o$. Conventionally, the scenario approach is solved with a large sample size to ensure that the original chance constraints are satisfied with a large $\eta$. However, as mentioned earlier, a large sample size in scenario approximation leads to overly conservative solutions. In contrast, we use optimization (\ref{cost_scenario})-(\ref{feasible_scenario}) as an intermediate step while constructing our desired distribution. Consequently, it is sufficient to solve (\ref{cost_scenario})-(\ref{feasible_scenario}) with a   low sample size (typically with $n_r=n_o \approx 10$).
%
%
%
%It is worth pointing out that the desired distribution  can always be constructed if we have access to sets $\mathcal{C}_{\widetilde{\textbf{w}}_{t-1}}$, $\mathcal{C}_{\widetilde{{_{j}^{o}}\boldsymbol{\xi}}_t}$. The construction of these two sets is guaranteed as long as we can obtain a feasible solution to the scenario approximation (\ref{cost_scenario})-(\ref{feasible_scenario}) for a small sample size.

\vspace{-0.33cm}
\subsection{GMM-KL Divergence (GMM-KLD) Based Approach}\label{kld_description}

%\begin{figure}
%    \centering
%    \includegraphics[width=1\linewidth]{results/kld-overview-v0.pdf}
%    \caption{Overview of our GMM-KL Divergence algorithm}
%    \label{fig:kld-overview}
%\end{figure}

% \noindent KL divergence is one of the most commonly used methods in statistics that gives a measure of how a given probability distribution is different from another. In other words, KL divergence is a valid choice for $\mathcal{L}_{dist}$. The primary challenge in using KL divergence for evaluating  $\mathcal{L}_{dist}$ is that we do not have access to the the probability densities associated with  $P_{f_t^j} (\textbf{u}_{t-1})$ and $P^{des}_{f_t^j}$. Furthermore, $P_{f_t^j} (\textbf{u}_{t-1})$ is parametrized in terms of control.  That is, for every $\textbf{u}_{t-1}\in \mathcal{C}$, we get a different distribution. In Algorithm \ref{algo:kld}, we present a workaround based control sampling and GMM fit.

\noindent KL divergence is extensively used to measure how a given probability distribution is different from another. In other words, KL divergence is a valid choice for $\mathcal{L}_{dist}$. The primary challenge in using KL divergence as a choice for  $\mathcal{L}_{dist}$ is that we do not have access to the the probability densities associated with  $P_{f_t^j} (\textbf{u}_{t-1})$ and $P^{des}_{f_t^j}$. To address this problem, we approximate $P_{f^j_t}$ and $P^{des}_{f_t^j}$ at any time $t$ using GMMs. It should be noted that  $P_{f_t^j} (\textbf{u}_{t-1})$ is parametrized in terms of control and changes for every $\textbf{u}_{t-1}\in \mathcal{C}$.

% \subsubsection{Overview of the method}
%The overview of the method is outlined in Fig. \ref{fig:kld-overview}. 

At a given time step, $t$, we first obtain samples from the desired distribution $P_{f_t^j}^{des}$ and subsequently fit a GMM over it. Then, for a given sampled control, we generate
different samples of $f^j_t(\textbf{w}_{t-1}^i, {_j^o}\boldsymbol{\xi}_{t}^j, \textbf{u}_{t-1})$ using the $(i, j)$ sample from $\textbf{w}_{t-1}$ and ${_j^o}\boldsymbol{\xi}_{t}$. We then fit GMM over it which now acts as an approximation of $P_{f_t^j}(\textbf{u}_{t-1})$ for the given sampled control. Using the GMM fit for $P_{f_t^j}^{des}$ and $P_{f_t^j}(\textbf{u}_{t-1})$  we can  then compute the KL divergence. This process is repeated for all the sampled control input and the one which minimizes $\mathcal{L}_{dist} + J(\textbf{u}_{t-1})$ is chosen as the solution of (\ref{cost_reform_gen})-(\ref{feas_reform_gen}).

\vspace{-0.35cm}
\subsection{RKHS based Approach}

\subsubsection{Formulating $\mathcal{L}_{dist}$ in terms of Moment Matching }
\noindent One of the vital building blocks of our approach based on RKHS embedding is the following theorem from \cite{moment_bound_lindsay}.

\small
\begin{theorem}\label{th_moment_bound}
 $\Vert P_{f_t^j}(\textbf{u}_{t-1})- P_{f_t^j}^{des}\Vert \leq B(d)$, $B(d)\rightarrow 0$, $d\rightarrow \infty$ 
\end{theorem}
\normalsize

\noindent where, $d$ refers to the order up to  which the moments of $P_{f_t^j}(\textbf{u}_{t-1})$ and $P_{f_t^j}^{des}$ are similar. The above theorem suggests that the difference between two distributions can be bounded by a non-negative function $B(d)$, which decreases with increasing order of moment $d$. The authors in \cite{moment_bound_lindsay} also show that this bound is particularly tight near the tail end of the distribution. Theorem \ref{th_moment_bound} provides a way of ensuring local similarity between two distributions through moment matching. In other words, moment matching is a valid choice for $\mathcal{L}_{dist}$

\newtheorem{remark}{Remark}

\vspace{-0.2cm}
\begin{remark}\label{local_similarity}
 For our purpose local similarity is sufficient since as we make the tail of  $P_{f_t^j}^{des}$ and  $P_{f_t^j}(\textbf{u}_{t-1})$ similar by matching higher order moments, we ensure that more and more of the mass of $P_{f_t^j}(\textbf{u}_{t-1})$ also gets shifted to the left of $f_t^j(.)=0$. This in turn leads to the satisfaction of chance constraints (\ref{chance_mpc}) with a higher $\eta$.  
\end{remark}

\vspace{-0.25cm}
\begin{remark}\label{gmm_rkhs_tuning}
The moment order $d$ can directly control the extent of local similarity and consequently acts as a surrogate for $\eta$. Higher $d$ leads to higher $\eta$ (see Fig \ref{fig:kld-vs-mmd-trajectories-o1}, \ref{fig:kld-vs-mmd-trajectories-o2}). Furthermore, $d$ only takes non-negative integer values. On the other hand, tuning GMM based solution is difficult as it relies on manipulating $\rho$ in (\ref{cost_reform_gen}) to prioritize minimizing MMD over primary cost $J(u)$. However, unlike $d$, the parameter $\rho$ does not have a statistical interpretation and thus it is difficult to ascertain whether a numerical change in $\rho$ will lead to any change in $\eta$ at all.
\end{remark}

%This is one of the critical differences between the KL divergence and RKHS based approach. The KL divergence would try to match the exact shape of $P_{f_t^j}^{des}$ and  $P_{f_t^j}(\textbf{u}_{t-1})$ while

\vspace{-0.15cm}
\subsubsection{ Moment Matching through RKHS Embedding} 
\noindent To the best of our knowledge, there is no mapping that directly quantifies the similarity between the first $d$ moments of two given distributions. However, a workaround can be devised based on the concept of embedding distributions in Reproducing Kernel Hilbert Space (RKHS) and Maximum Mean Discrepancy (MMD) distance. The key ideas here are based on the results from \cite{scholkopf}, \cite{scholkopf2}.

Let $\mu_{P_{f_t^j}}, \mu_{P_{f_t^j}^{des}}$ represent the RKHS embedding of $P_{f_t^j}$, $P_{f_t^j}^{des}$ given by the following equations:

\vspace{-0.35cm}
\small
\begin{equation}
\mu_{P_{f_t^j}}(\textbf{u}_{t-1}) = \sum_{p=1}^{p=n}\sum_{q=1}^{q=n}\alpha_p\beta_qk(f_t^j(\textbf{w}_{t-1}^p, {_{j}^{o}}\boldsymbol{\xi}_t^q,\textbf{u}_{t-1}),.)
\label{rkhs_chance2}
\end{equation}
\normalsize

\vspace{-0.3cm}
\small
\begin{equation}
\mu_{P_{f_t^j}^{des}} = \sum_{p=1}^{p=n_r}\sum_{q=1}^{q=n_o}\lambda_p\varphi_qk(f_t^j(\widetilde{\textbf{w}}_{t-1}^p, \widetilde{{_{j}^{o}}\boldsymbol{\xi}}_t^q, \textbf{u}_{t-1}^{nom}),.)
\label{rkhs_chance1_desired},
\end{equation}
\normalsize

\vspace{-0.1cm}

\noindent where, $k(.,.) : \Re^n \times \Re^N \rightarrow \Re$ is a positive definite function called the kernel. To exploit Theorem \ref{th_poly_mmd}, we use the polynomial kernel of order $d$  defined in (\ref{poly_ker}) defined for some arbitrary vectors $\textbf{x}_1, \textbf{x}_2$.

% \vspace{-0.2cm}
%\small
%\begin{eqnarray}
%{k}(f_t^j,f_t^j) = {(f_t^j(\widetilde{\textbf{w}}_{t-1}^p, \widetilde{{_{j}^{o}}\boldsymbol{\xi}}_t^q,\textbf{u}_{t-1})*f_t^j(\widetilde{\textbf{w}}_{t-1}^p, \widetilde{{_{j}^{o}}\boldsymbol{\xi}}_t^q,\textbf{u}_{t-1})+c)}^d \label{poly_ker_defn}\\ 
%c>0,d>0, c\in \mathcal{R}, d\in \mathcal{N} \label{poly_ker_param}
%\end{eqnarray}

\vspace{-0.4cm}
\small
\begin{equation}
k(\textbf{x}_1, \textbf{x}_2) = (1+\textbf{x}_1^T\textbf{x}_2)^d.
\label{poly_ker}
\end{equation}
\normalsize

\vspace{-0.25cm}

\noindent Note that the $d$ in (\ref{poly_ker}) is same as that used in Theorem \ref{th_moment_bound}.
The constants $\alpha_p, \beta_q, \lambda_p, \varphi_q$ play a vital role in the RKHS embedding, and we discuss them towards the end of this section. Now, consider the following theorem based on \cite{scholkopf2}, (\cite{scholkopf4}, pp-15) assuming that the RKHS embedding is constructed through polynomial kernel of order $d$.

\small
\begin{theorem}\label{th_poly_mmd}
If $\overbrace{\Vert \mu_{P_{f_t^j}}(\textbf{u}_{t-1})-\mu_{P^{des}_{f_t^j}} \Vert}^{MMD} \rightarrow 0$, then moments of $P_{f_t^j}$  and  $P_{f_t^j}^{des}$ up to order $d$ become similar.
\end{theorem}
\normalsize

\noindent That is, decreasing the residual of MMD distance becomes a way of matching the first $d$ moments of the distribution $P_{f_t^j}$ and $P_{f_t^j}^{des}$. Theorem  \ref{th_poly_mmd} allows us to make the following choice

\vspace{-0.25cm}
\small
\begin{equation}
\mathcal{L}_{dist} = \Vert \mu_{P_{f_t^j}}(\textbf{u}_{t-1})-\mu_{P^{des}_{f_t^j}} \Vert.
\label{l_dist_mmd}
\end{equation}
\normalsize

\vspace{-0.2cm}
\noindent The MMD (\ref{l_dist_mmd}) can be expanded in the following manner

\vspace{-0.35cm}
\small
\begin{eqnarray}\nonumber
\Vert \mu_{P_{f_t^j}}(\textbf{u}_{t-1})-\mu_{P^{des}_{f_t^j}}\Vert^2 = \langle \mu_{P_{f_t^j}}(\textbf{u}_{t-1}),\mu_{P_{f_t^j}}(\textbf{u}_{t-1})\rangle\\-2\langle \mu_{P_{f_t^j}}(\textbf{u}_{t-1}),\mu_{P^{des}_{f_t^j}}\rangle+\langle \mu_{P^{des}_{f_t^j}}, \mu_{P^{des}_{f_t^j}}\rangle \label{kernel_trick1}
\end{eqnarray}
\normalsize
\vspace{-0.25cm}

For a given $\textbf{u}_{t-1}$, each of the terms in (\ref{kernel_trick1}) is an inner product of two functions that are linear combinations of kernel functions and thus can be easily simplified using the so-called "kernel trick". These inner products give rise to large matrices, and hence easily parallelized using GPU.

\vspace{-0.35cm}
\subsection{Reduced Set Methods}\label{reduced_set}
\noindent Let $\hat{\textbf{w}}_{t-1}^1, \hat{\textbf{w}}_{t-1}^2 \dots \hat{\textbf{w}}_{t-1}^N$ and $\hat{{_{j}^{o}\boldsymbol{\xi}}}_t^1, \hat{{_{j}^{o}\boldsymbol{\xi}}}_t^2 \dots \hat{{_{j}^{o}\boldsymbol{\xi}}}_t^N$  represent $N$ i.i.d samples of $\textbf{w}_{t-1}$, ${_{j}^{o}\boldsymbol{\xi}}_t$ respectively. Further, let  $\textbf{w}_{t-1}^1, \textbf{w}_{t-1}^2 \dots \textbf{w}_{t-1}^n$ and ${_{j}^{o}\boldsymbol{\xi}}_t^1, {_{j}^{o}\boldsymbol{\xi}}_t^2 \dots {_{j}^{o}\boldsymbol{\xi}}_t^n$ represent a subset (reduced set) of the i.i.d samples. It is implied that $n<<N$. Now, intuitively, a reduced set method would re-weight the importance of each sample from the reduced set such that it would  retain as much  information as possible from the original i.i.d samples. The weights $\alpha_p$ associated with $\textbf{w}_{t-1}$   are computed through the following optimization and are then used to compute the RKHS embedding in (\ref{rkhs_chance2}). The same process can also be used to compute $\beta_q, \lambda_p, \varphi_q$.

\vspace{-0.45cm}
\small
\begin{equation}
\alpha_p = \arg\min_{\alpha_p}\Vert \frac{1}{N}\sum_{i=1}^{i=N}k(\hat{\textbf{w}}_{t-1}^i,.)-\sum_{p=1}^{p=n}\alpha_pk(\textbf{w}_{t-1}^p,.)\Vert_2,\sum \alpha_p =1 
\label{eq:reduced-set}
\end{equation}
\vspace{-0.5cm}
\normalsize

%An example of these are detailed out in \cite{GMM-fit-quality},\cite{GMM-fit-quality1} that talk about the success of GMM fits based on various lower bounds on sample sizes and other GMM parameters. 

% \vspace{-0.1cm}

\vspace{-0.5cm}
\subsection{Performance Guarantees}\label{sec:performance-guarantees}
\noindent Both our GMM-KLD and RKHS based approaches work with only sample-level information without assuming any parametric form for the underlying distribution. Thus, the performance guarantees on collision avoidance depend on the following aspects. First, on how well are we modeling the distribution of our collision avoidance function (PVO) for a finite sample size. Second, on whether our modeling improves as the sample size increase: a property popularly known as consistency in estimation. Finally, on whether we can tune our model to produce diverse trajectories with different probability of avoidance. Remark \ref{gmm_rkhs_tuning} already addresses the third question. Moreover, the first two questions about GMM fit of distributions have already been established in the existing literature \cite{GMM-fit-quality}, \cite{GMM-fit-quality1}. Thus, in this subsection, we focus on the first two questions regarding our RKHS based approach.

\noindent To show the consistency of (\ref{rkhs_chance2}), we compute a ground truth embedding in the following manner:

\small
\vspace{-0.4cm}
\begin{equation}
\overline{\mu_f(\textbf{w}_{t-1},{_j^o}\boldsymbol{\xi}_{t},\textbf{u}_{t-1})}=(1/l^2)\sum_{i=1}^{l}\sum_{j=1}^{l}k(f({\textbf{w}_{t-1}}^i,{{_j^o}\boldsymbol{\xi}_{t}}^j,\textbf{u}_{t-1}),.),
\label{rkhs_gt}
\end{equation}
\normalsize

\noindent where, $\overline{\mu_f(\textbf{w}_{t-1},{_j^o}\boldsymbol{\xi}_{t},\textbf{u}_{t-1})}$ is similar to $\mu_{P_{f_t^j}}(\textbf{u}_{t-1})$ except that the former is computed over a larger sample size $l$. That is, $l>>n$. We can analyze the consistency by constructing the following error function from \cite{scholkopf},\cite{scholkopf2} for a fixed value of $\textbf{u}_{t-1}$.

\begin{figure}[H]
    \centering
    \vspace{-0.2cm}
    \includegraphics[width=0.3\textwidth]{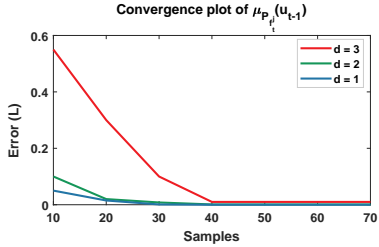}
    \vspace{-0.2cm}
    \caption{We illustrate the convergence of (\ref{error_measure}) with increasing sample size $n$ and hence show that $\mu_{P_{f_t^j}}(\textbf{u}_{t-1})$ is a consistent estimator. We evaluated equation (\ref{error_measure}) for a polynomial kernel, with different orders of moment $d=1,2,3$.}
    \vspace{-0.33cm}
    \label{fig:rkhs-convergence} 
\end{figure}

\small
\vspace{-0.5cm}
\begin{align}
L=\Vert\mu_f(\textbf{w}_{t-1},{_j^o}\boldsymbol{\xi}_{t},\textbf{u}_{t-1})-\overline{\mu_f(\textbf{w}_{t-1},{_j^o}\boldsymbol{\xi}_{t},\textbf{u}_{t-1})}\Vert_2^2,
\label{error_measure}
\end{align}
\normalsize
\vspace{-0.5cm}

\noindent and analyzing its behavior for increasing value of $n$. The results are summarized in  Fig.\ref{fig:rkhs-convergence}. As shown, for moment order $d=1,2,3$ in the polynomial kernel function, we get very low error for a sample size as low as 40.
%The simulations presented in the paper were obtained with $d=1,2,3$ and $50$ samples of $\textbf{w}_{t-1}$ and ${_j^o}\boldsymbol{\xi}_{t}$ each.}

%The third aspect of tuning the model to produce diverse trajectories, with varied $\eta$ is detailed out in the simulations and validation section, followed by additional results in the supplementary material\footnote{The supplementary material with additional details can be found at  \href{https://robotics.iiit.ac.in/uploads/Main/Publications/rkhs-collision-avoidance}{https://robotics.iiit.ac.in/uploads/Main/Publications/rkhs-collision-avoidance}}.

% \vspace{-0.3cm}

\section{Simulation and Validation}\label{sec:simulation-validation}
Additional  results, realtime experiments  and  detailed  derivation  of the kernel matrices along with the implementation details are provided in the supplimentary material\footnote{\href{https://robotics.iiit.ac.in/uploads/Main/Publications/rkhs-collision-avoidance/}{robotics.iiit.ac.in/uploads/Main/Publications/rkhs-collision-avoidance}}.
% We used our RKHS based algorithm to simulate the navigation of quadrotor in the X-Y plane amongst dynamic obstacles. The motion model was approximated as a double integrator system, wherein acceleration control inputs are subjected to random perturbations. To keep the simulation realistic, we estimated the motion and perception noise associated with a real quadrotor. For this, we ran a quadrotor in open loop along several trajectories multiple number of times. We then used the odometry readings and the ground truth to compute the error distribution for the positions, velocities and accelerations. The results obtained for one trajectory are shown in Fig.\ref{groundtruth_exp}-\ref{perception_uncert}. As can be seen, the error distributions indeed show a non-Gaussian trend. We fit a Pearson distribution with moments upto fourth order to the error statistics. Note that the fitted Pearson distribution will be in a black-box form and thus, its parametric form cannot be known. The uncertainty in state and control were then simulated as nominal values perturbed with random errors drawn from the fitted Pearson distribution. A more detailed description is presented in the supplementary material\footnote{The supplementary material with additional details and derivations can be found in \url{https://robotics.iiit.ac.in/uploads/Main/Publications/Bharath_et_al_icra19/}}.

\begin{figure*}[t]
    \centering
    % \vspace{5cm}
    \includegraphics[width=0.9\textwidth]{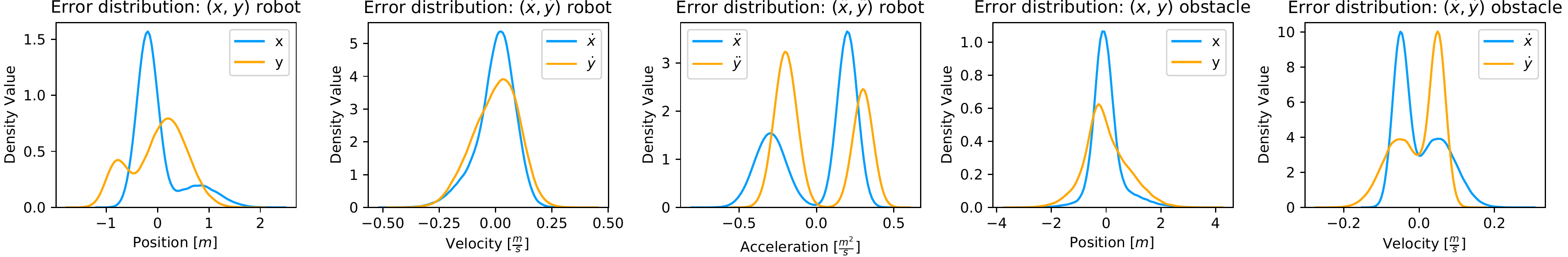}
    \vspace{-0.25cm}
    \caption{Figures show the error distributions for position, velocity, acceleration of the robot and position, velocity of the obstacle.}
    % \caption{Figures show the error distributions for position, velocity, acceleration of the robot and position, velocity of the obstacle. The figures also show the Gaussian approximation of the actual non Gaussian error statistics.}
    \vspace{-0.25cm}
    \label{fig:error-distributions}
\end{figure*}

We used our RKHS based algorithm to simulate the navigation of quadrotor in the X-Y plane amongst dynamic obstacles. We approximate the motion model as a double integrator system, wherein we subject the acceleration control inputs to random perturbations. To showcase the potential of our algorithm, we used non-Gaussian error distributions for the positions, velocities, and accelerations of the robot and obstacle as applicable. Fig. \ref{fig:error-distributions} illustrates the error distributions used. 50 samples of $\textbf{w}_{t-1}$  and ${_j^o}\boldsymbol{\xi}_{t}$ were used to evaluate both RKHS and GMM based approach. We also used a set of 50 control samples to search for an optimal solution of (\ref{cost_reform_gen})-(\ref{feas_reform_gen}).

\subsubsection{Evaluation of the RKHS based Approach for one obstacle benchmark}\label{sec:one-obs-benchmark}
The solution process and results are summarized in the results shown in the Fig. \ref{fig:kld-vs-gmm-one-obstacle-results}. As described previously, the solution process starts with the construction of the desired distribution $P^{des}_{f^j_t}$. Subsequently, we ensure that the distribution of $P_{f^j_t}(\textbf{u}_{t-1})$ is similar to $P^{des}_{f^j_t}$ (atleast near the tail end) by choosing an appropriate $\textbf{u}_{t-1}$ and degree of the polynomial kernel $d$. This is highlighted in Fig. \ref{fig:kld-vs-mmd-distributions-o1}, which plots the distribution of $P_{f_t^j}(\textbf{u}_{t-1}) $\footnote{Kernel density estimators are used to plot the distributions.} for different degrees, $d$, of the polynomial kernel. It is clearly shown that as $d$ increases, the distributions $P^{des}_{f^j_t}$ and $P_{f^j_t}(\textbf{u}_{t-1})$ become more alike (at least near the tail end). 
In this figure $P^{des}_{f^j_t}$ is highlighted in a grey shade, while the PDF of $P_{f_t^j}(\textbf{u}_{t-1})$ for $d=1,2,3$ are outlined in blue, orange and green respectively.
The increase in the similarity between these two distributions can be directly correlated with the actual collision avoidance shown in Fig. \ref{fig:kld-vs-mmd-trajectories-o1}. An increased value of $d$ results in lesser overlap between the samples of the robot and obstacle, thus resulting in higher value of $\eta$. Fig. \ref{fig:kld-vs-mmd-costs-o1} shows that the control cost  ($\Vert \textbf{u}_{t-1} \Vert^2_2$) and tracking error ($\Vert\boldsymbol{\overline{\xi}}_{t} - \boldsymbol{\xi}_t^d\Vert_2^2$) increases at higher values of $d$, thus making maneuvers more conservative.  The top row of figure \ref{fig:kld-vs-mmd-trajectories-o1}, shows progressively conservative avoidance with increase in $d$, while the bottom row shows that the progress to the goal is slowed with increase in $d$, due to such conservative maneuvers. The maneuver for $d=3$, resembles closely the maneuver due to GMM-KLD.

\subsubsection{Evaluation of the GMM-KL Divergence Approach}
\begin{figure*}[h]
     \centering
     \subfigure[Figures show snapshots of collision avoidance maneuvers obtained with RKHS (first three columns) and GMM-KLD (last column) based approaches. The top row shows that the overlap between robot and obstacle uncertainty reduces as $d$ increases resulting in collision avoidance with higher $\eta$. For $d =3$, the maneuver obtained with RKHS is similar to that obtained with GMM-KLD approach. The bottom row shows that the progress in the trajectory towards the goal is slower as $d$ increases. For example, after 3.5s, for $d=1$, the robot has already reached the goal. In contrast, for $d = 3$, it has just maneuvered the obstacle. A very similar slow progress to the goal is visible in GMM maneuver. ]{
        \centering
        \includegraphics[scale=0.5,width=0.96\textwidth]{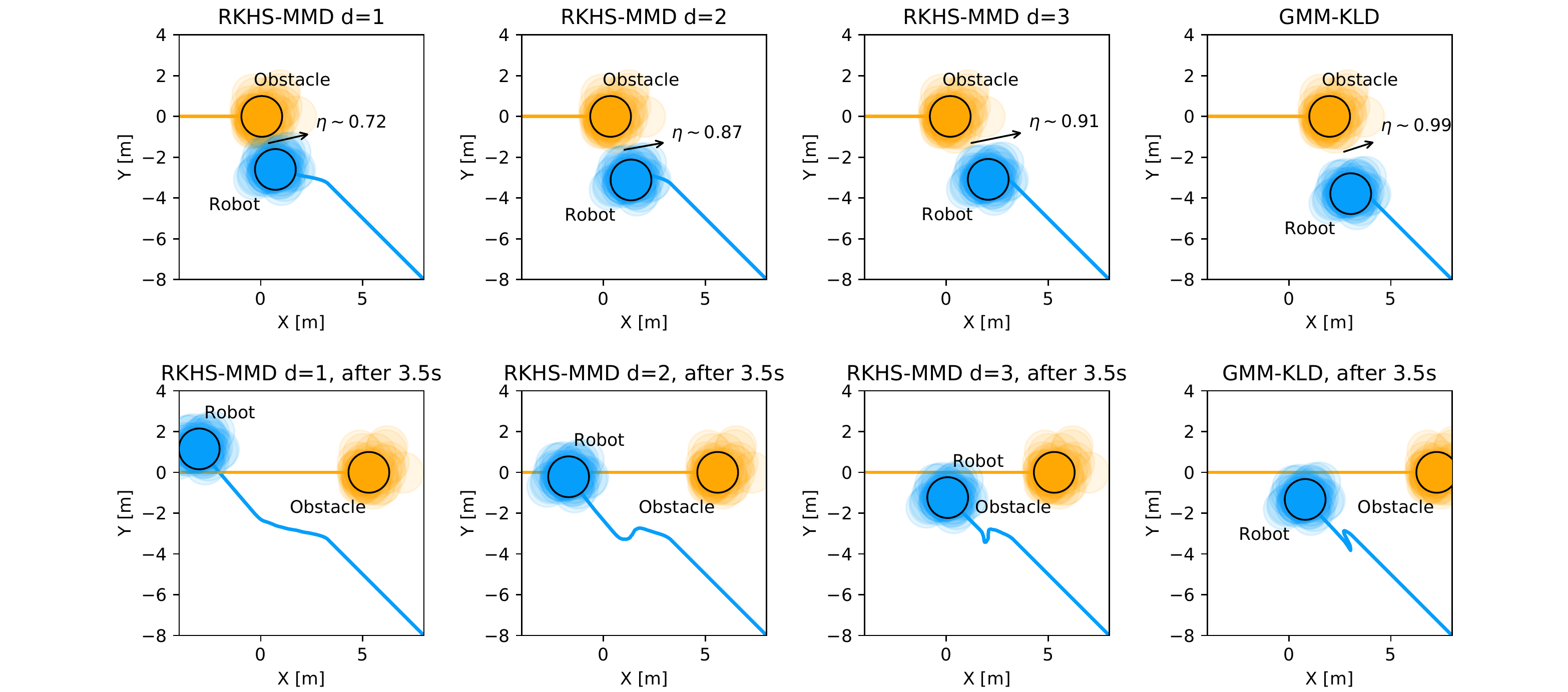}
        \label{fig:kld-vs-mmd-trajectories-o1}
     }
     \subfigure[Lower L2 norm signifies less jerky motion of the robot. The cumulative velocity tracking cost quantifies how deviant the velocity of the robot was from the desired velocity (the desired velocity is the velocity towards the goal). The tracking cost can also be related to the time taken to reach the goal.]{
        \centering
         \includegraphics[width=0.52\textwidth]{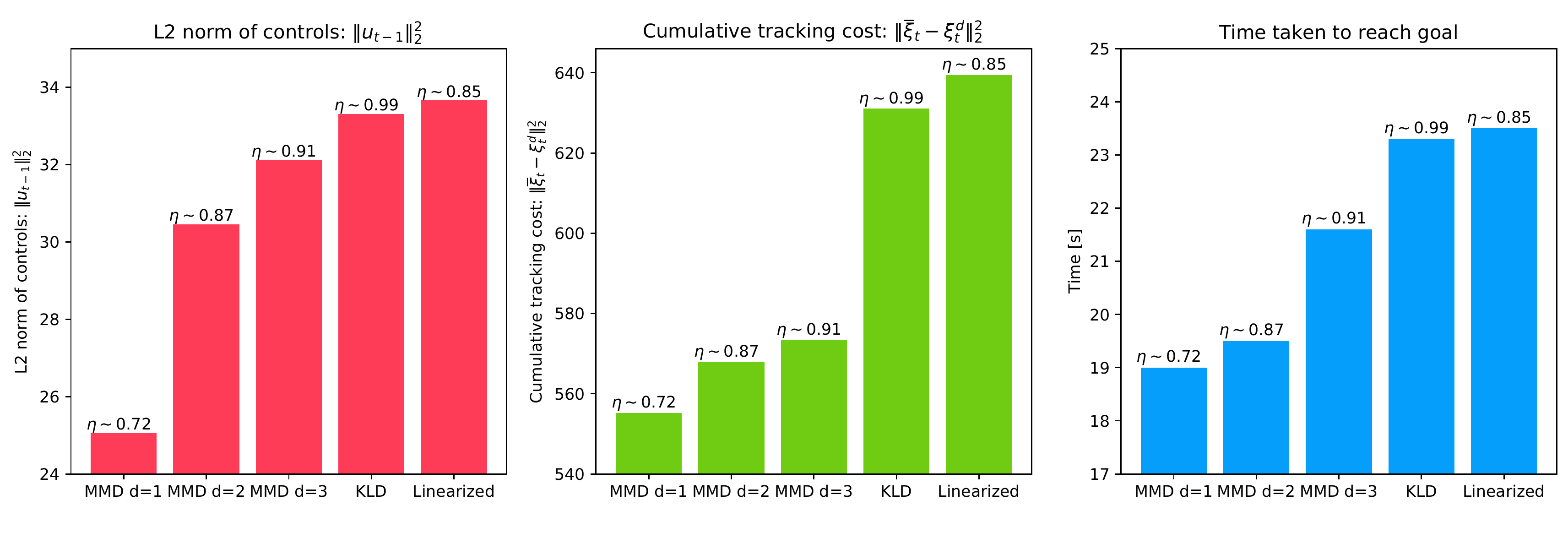}
         \label{fig:kld-vs-mmd-costs-o1}
     }
     \hfill
     \subfigure[Distribution of velocity obstacle $P_{f^j_t}(\textbf{u}_{t-1})$ obtained after solving the RKHS based approach for different values of $d$. The distribution obtained with GMM-KLD based approach is also shown. ]{
         \centering
         \vspace{-0.3cm}
         \includegraphics[width=0.38\textwidth]{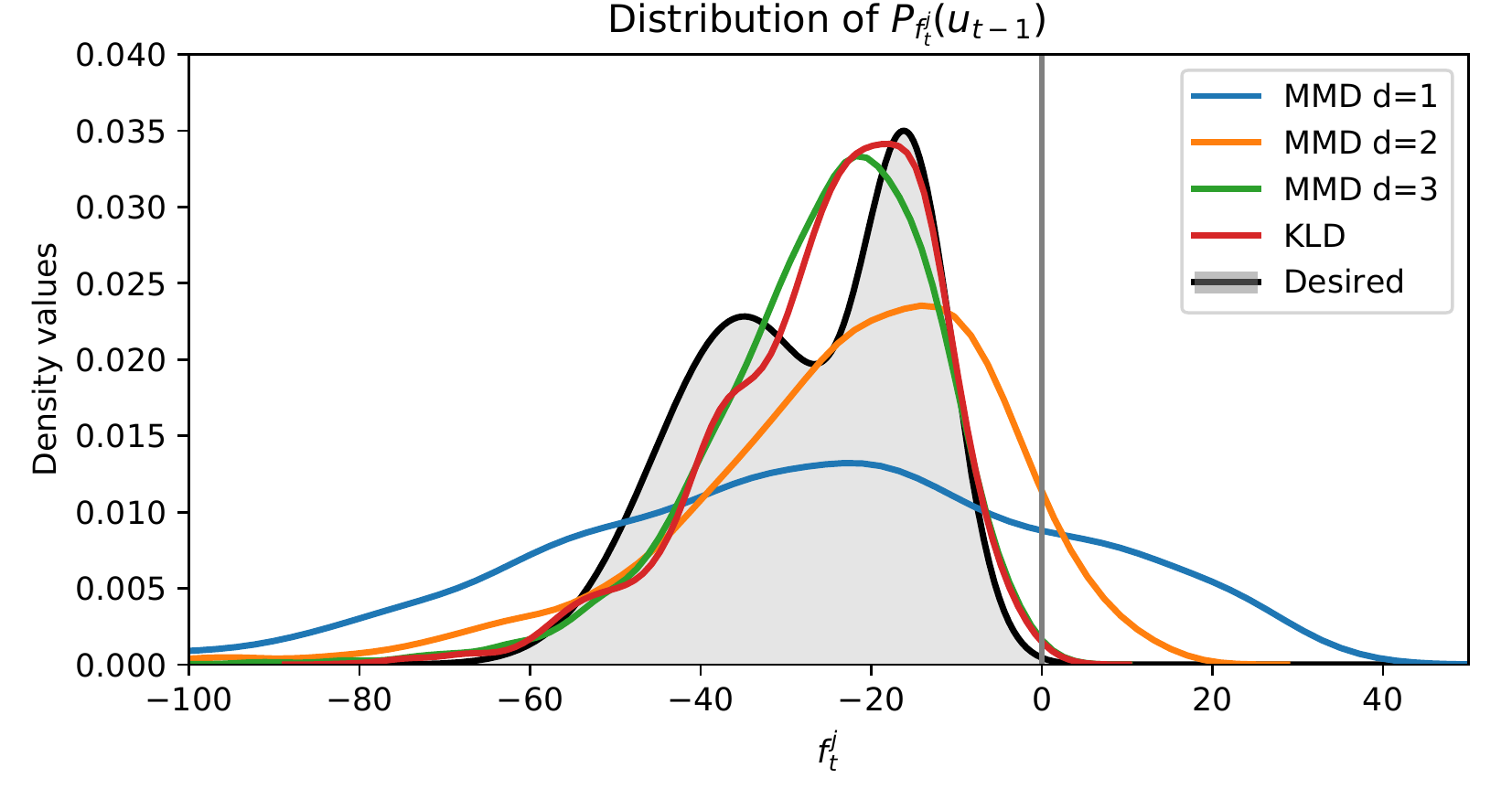}
         \label{fig:kld-vs-mmd-distributions-o1}
     }
    \vspace{-0.2cm}
    \caption{Simulation results for one robot and one obstacle scenario.}
    \label{fig:kld-vs-gmm-one-obstacle-results}
    \vspace{-0.5cm}
\end{figure*}

For the same one obstacle benchmark, we evaluate the GMM-KL Divergence approach according to section \ref{kld_description}. In short, we select the control, $\textbf{u}_{t-1}$, that minimizes the cost function given in (\ref{cost_mpc}) by choosing $\mathcal{L}_{dist}$ as KL divergence. The results are shown in Fig. \ref{fig:kld-vs-mmd-distributions-o1}. It is evident that $P_{f^j_t}(\textbf{u}_{t-1})$ matches the shape of $P^{des}_{f^j_t}$, thus satisfying one of the basic objectives of minimizing a KL Divergence scheme.  However, as mentioned in Remark \ref{gmm_rkhs_tuning}, it is difficult to tune the GMM-KLD approach to produce diverse trajectories. For this specific result, we obtained a solution for which $\eta\approx0.99$. Thus, the trajectories have a high safety factor but at the same time have higher tracking error and control cost.
%
%matching $P^{des}_{f^j_t}$ through a KL Divergence scheme results in maneuvers that have an increased $\eta$ but are highly conservative. This can be observed in Fig. \ref{fig:kld-vs-mmd-trajectories-o1} where the snapshot titled "GMM-KLD" has less or no overlap between the samples of the robot and obstacle. We observed 

\vspace{-0.45cm}
\subsection{Two obstacle benchmark}
\begin{figure*}[h]
     \centering
     \subfigure[The first three columns represent the different paths taken by the robot upon using an RKHS based approach with increasing values of $d$. The last column represents the path obtained with GMM-KLD based approach.]{
        \centering
        \includegraphics[width=0.96\textwidth]{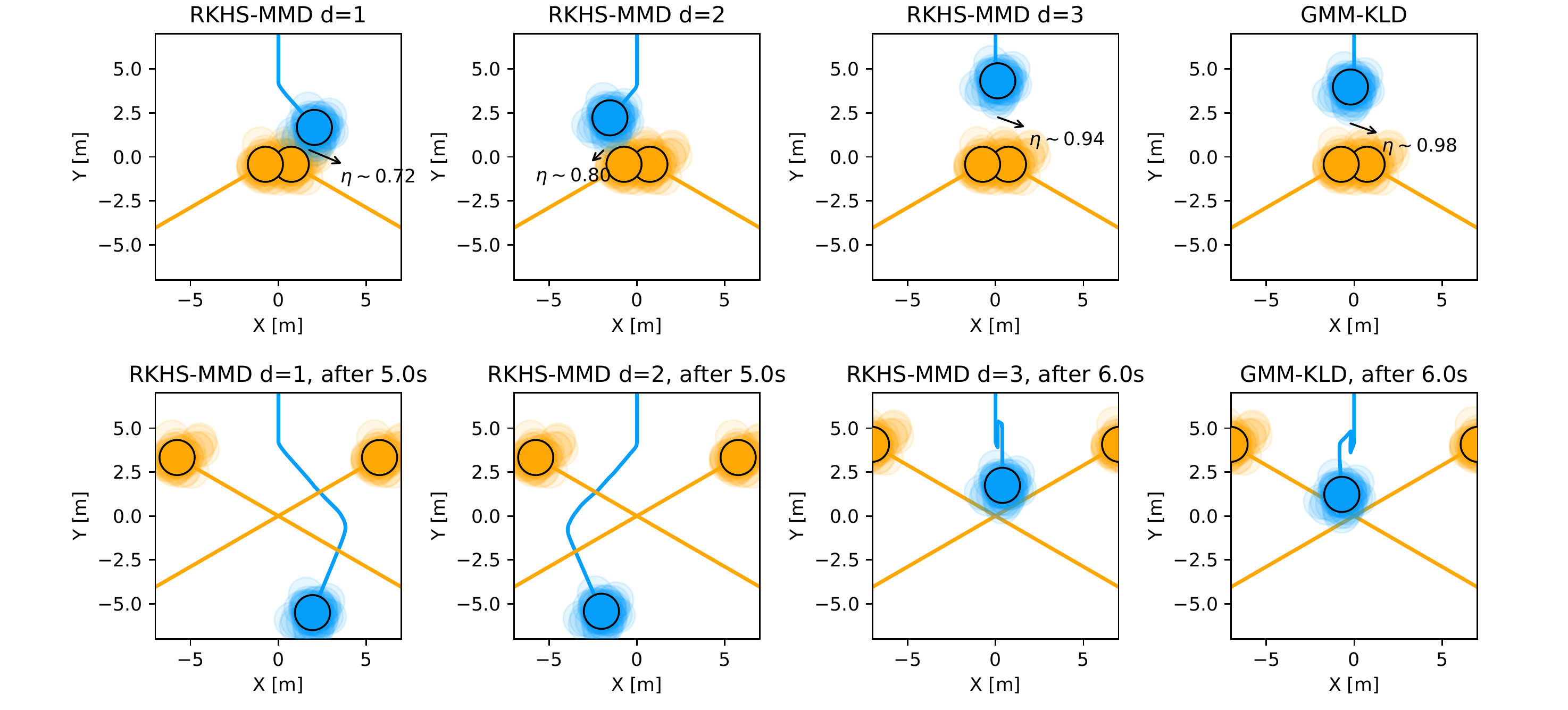}
    \label{fig:kld-vs-mmd-trajectories-o2}
     }
     \subfigure[Comparison of L2 norm of controls, desired velocity tracking cost and time taken to reach goal for the one robot-two obstacle benchmark.]{
        \centering
         \includegraphics[width=0.6\textwidth]{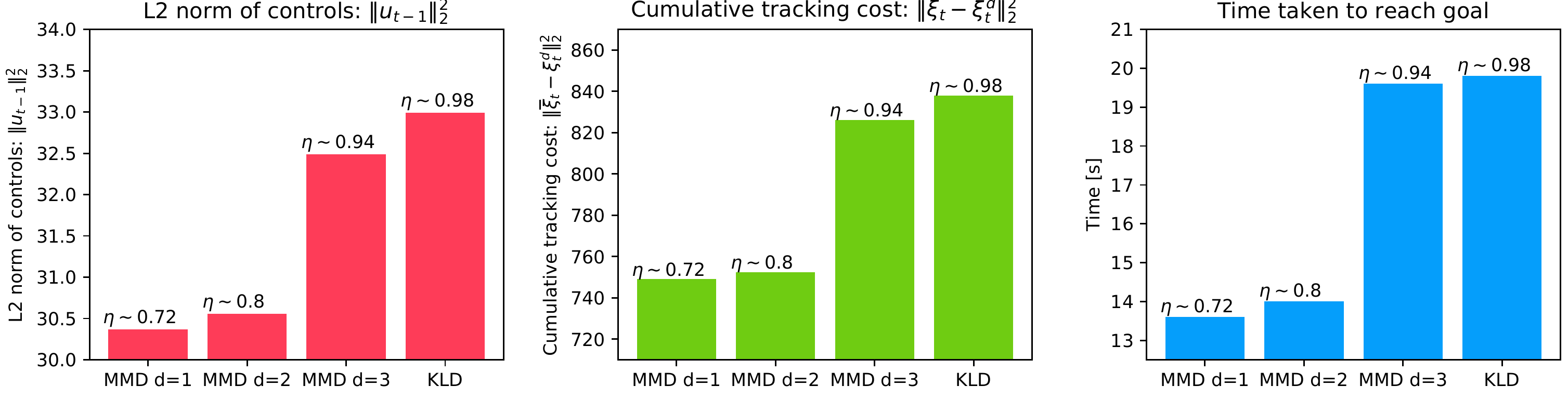}
         \label{fig:kld-vs-mmd-costs-o2}
     }
     \hfill
    %  \subfigure[Collision cone, $P_{f^j_t}$, distributions for different degrees of the kernels used (RKHS based) and KL divergence.]{
    %      \centering
    %      \includegraphics[width=0.34\textwidth]{results/dists-small-v1.pdf}
    %      \label{fig:kld-vs-mmd-distributions}
    %  }
        \subfigure[Time complexity comparison for RKHS-MMD and GMM-KL Divergence based approaches for different number of obstacles.]{
            \centering
            \vspace{-2cm}
            \includegraphics[width=0.35\linewidth]{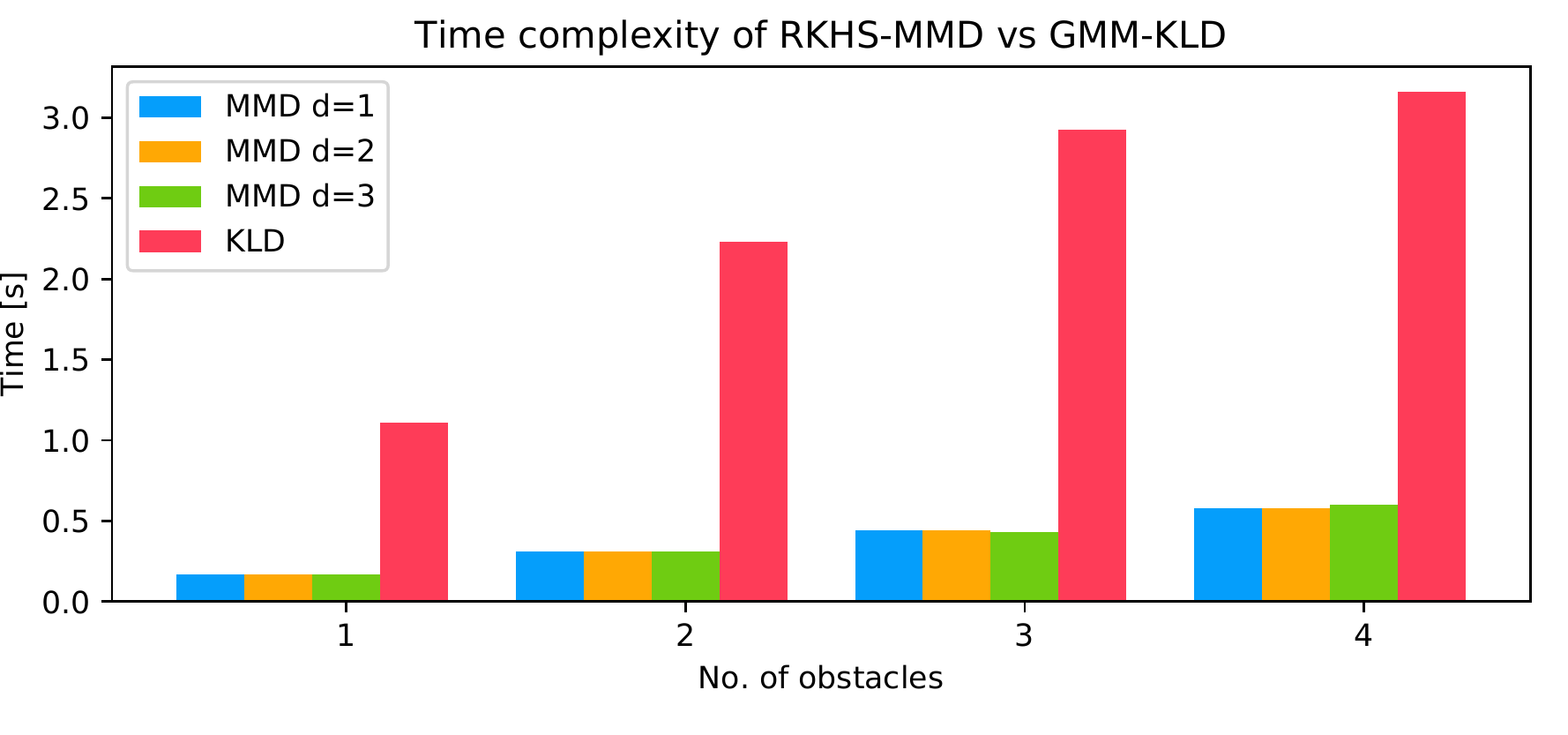}
            \label{fig:kld-vs-mmd-time-complexity}
        }
        
        \caption{Simulation results for one robot and two obstacles scenario.}
        \vspace{-0.5cm}
        \label{fig:kld-vs-mmd-results-o2}
\end{figure*}

Here we extend the proposed schemes to a two obstacle benchmark, where similar trends as obtained in section \ref{sec:one-obs-benchmark} are observed. Fig. \ref{fig:kld-vs-mmd-trajectories-o2} shows that an increased value of $d$ results in higher value of $\eta$ but at the same time leads to higher tracking errors ($\Vert\boldsymbol{\overline{\xi}}_{t} - \boldsymbol{\xi}_t^d\Vert_2^2$ and control costs $\Vert \textbf{u}_{t-1} \Vert^2_2$ (refer Fig. \ref{fig:kld-vs-mmd-costs-o2}).

\vspace{-0.45cm}
\subsection{Comparative results based on Computational Time}

The table \ref{tab:kld-mmd-1obs-time} shows the computation time for RKHS and GMM-KLD based approaches on an Intel i7,4770 laptop. The GMM-KLD approach takes a computational time of 1.24s, while the RKHS based approach takes 0.17s to achieve a similar $\eta$. The reduced computation time of the later stems primarily from two reasons. (i) Unlike GMM-KLD approach, RKHS based approach avoids the need to first fit a parametric form to the distribution of PVO and the desired distribution and then compute collision avoidance control. (ii) Furthermore, we exploit the kernel trick to reduce the MMD computation in it to just matrix multiplication, which can be efficiently parallelized on GPUs (NVIDIA GTX 1050 in our case). Fig. \ref{fig:kld-vs-mmd-time-complexity} shows the difference in computation time between the two approaches increases with the number of obstacles. The time to compute maneuvers grows faster for the GMM-KLD approach with an increase in the number of obstacles, so much so for four obstacles, the GMM approach is 2.5s slower than the RKHS approach. Whereas for one obstacle scenario GMM is 1s slower.

\vspace{-0.4cm}
\subsection{Experimental Results}
We implemented the RKHS framework on a Bebop drone equipped with a monocular camera. A person walking with an April tag marker in front of him constitutes the moving object. We compute the distance, velocity, and bearing to the marker using the April Tag library. We obtain the state and velocity of the drone from the onboard odometry. The state, velocity, control, and perception/measurement noise distributions were computed through a non-parametric Pearson fit over experimental data obtained over multiple runs. 

\begin{figure*}[h]
    \centering
    \subfigure{
        \centering
        \includegraphics[width=0.3\textwidth]{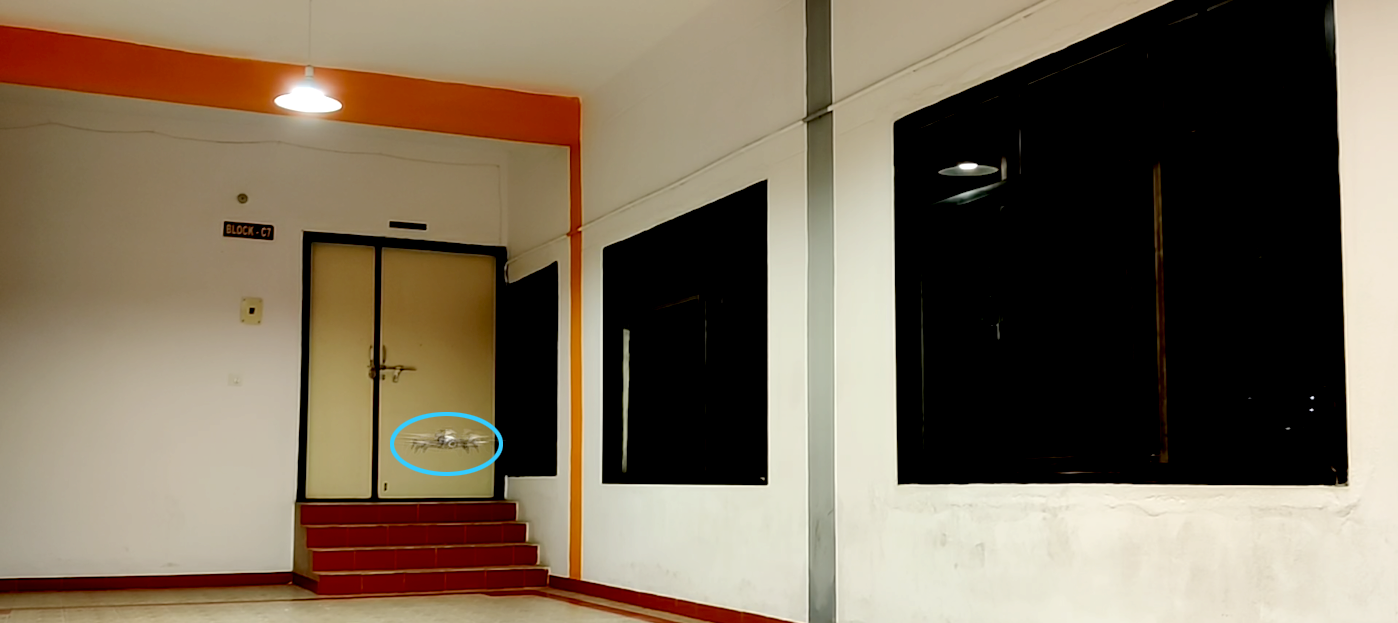}
    }
    \subfigure{
        \centering
        \includegraphics[width=0.3\textwidth]{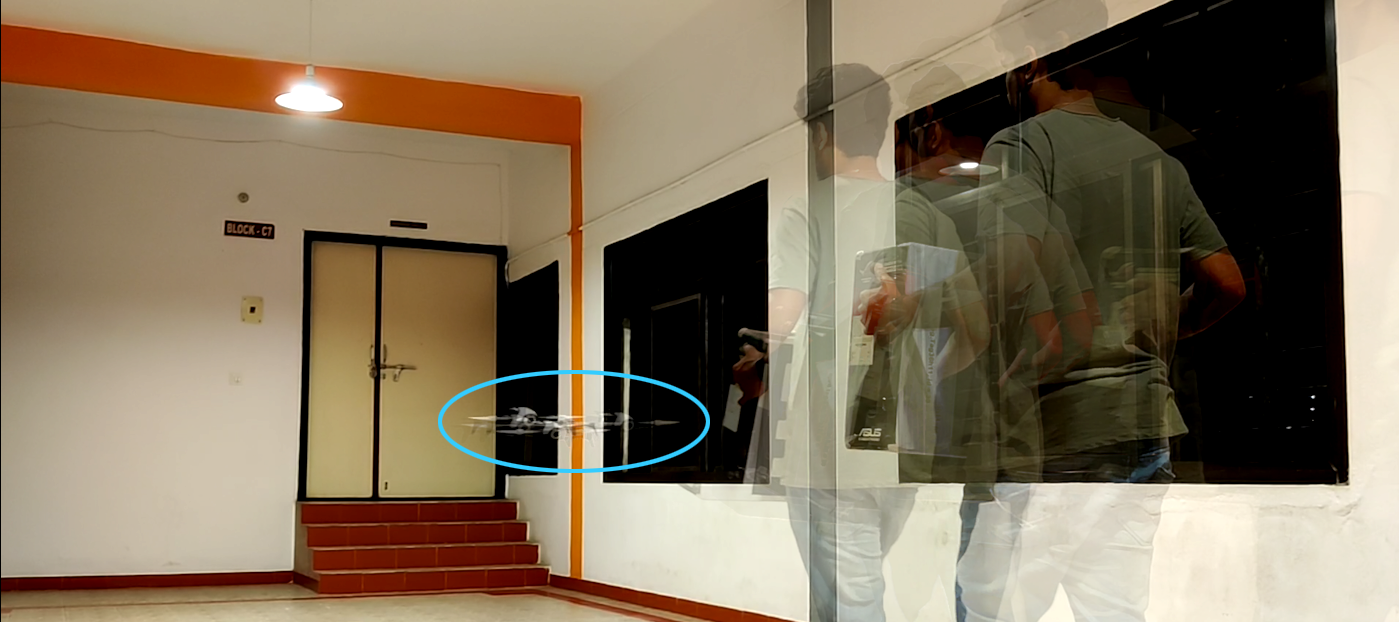}
    }
    \subfigure{
        \centering
        \includegraphics[width=0.3\textwidth]{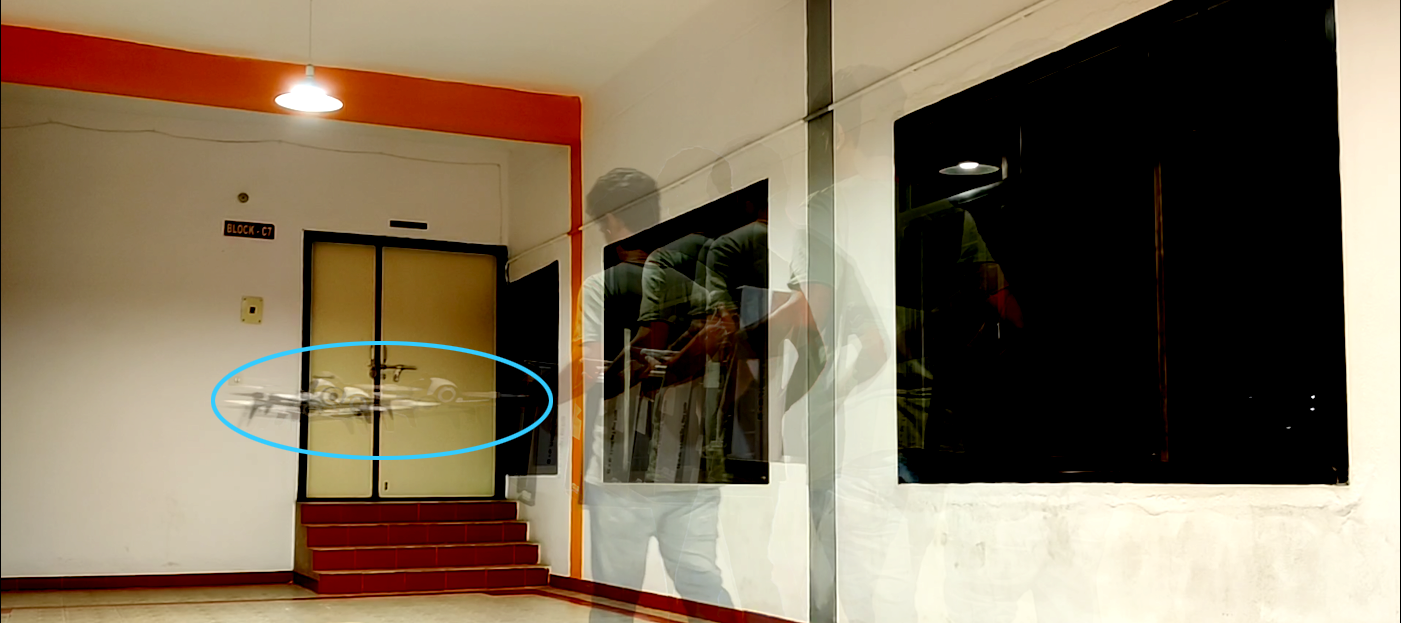}
    }
    \vspace{-0.2cm}
    \caption{Snapshots of the real-time collision avoidance maneuver. The snapshots show the instances of the robot sample from a non-parametric Pearson fit over experimental data. The same applies to the obstacle (human moving). Throughout the experiment, an approximate confidence ($\eta$) of 0.75 is maintained.}
    \vspace{-0.25cm}
    \label{fig:real-run-snaps}
\end{figure*}

We performed 15 experimental runs to evaluate the RKHS method, which successfully avoided the obstacle on 75\% of the 15 runs. Whereas naive deterministic maneuvers avoided the moving target, only 40\% of the 15 runs. Fig. \ref{fig:real-run-snaps} shows the snapshots of the experimental runs.

\begin{table}[]
\scriptsize
\caption{Time complexity for single obstacle}
\vspace{-0.3cm}
\begin{center}
\label{tab:kld-mmd-1obs-time}
\begin{tabular}{|c|c|c|c|c|}
\hline
\multicolumn{2}{|c|}{RKHS-Approach} & \multicolumn{3}{c|}{GMM-KLD} \\ \hline
\multirow{2}{*}{d} & \multirow{2}{*}{Time} & \multirow{2}{*}{\begin{tabular}[c]{@{}c@{}}Time for\\ GMM\end{tabular}} & \multirow{2}{*}{\begin{tabular}[c]{@{}c@{}}Time for\\ KLD\end{tabular}} & \multirow{2}{*}{\begin{tabular}[c]{@{}c@{}}Total\\ Time\end{tabular}} \\
 &  &  &  &  \\ \hline
1 & 0.17s & \multirow{3}{*}{0.59s} & \multirow{3}{*}{0.644s} & \multirow{3}{*}{1.24s} \\
2 & 0.17s &  &  &  \\
3 & 0.17s &  &  &  \\ \hline
\end{tabular}
\end{center}
\normalsize
\vspace{-0.7cm}
\end{table}

\vspace{-0.4cm}
%\input{chapters/limitations.tex}
%\vspace{-0.4cm}
%\input{chapters/multi-agent.tex}
% \vspace{-0.35cm}
\section{Conclusions and Future Work}
\vspace{-0.15cm}
In this paper, we formulated a robust MPC/CCO as a problem of distribution matching. We illustrated two approaches that give tractable solutions to optimization problem with complex chance constraints like PVO under non-parametric uncertainty. We first proposed a baseline method that approximates the distribution of the chance constraints with a \textbf{GMM} model and then proceed to perform distribution matching using \textbf{KL} divergence. Our second method is built on the possibility of embedding distributions in the Reproducing Kernel Hilbert Space (\textbf{RKHS}). We evaluated both the GMM-KLD and the RKHS based approaches for quality of maneuvers and computational time. The RKHS based approach results in lower tracking errors and control costs than the GMM-KLD based approach. Another distinct advantage of the former is the massive reduction in computational time as compared to the latter. To the best of the author's knowledge, this is a first such work that brings the benefits of RKHS embedding to the domain of robust MPC/CCO. This has enormous potential benefits for computationally efficient motion planning and control under uncertainty.

% \vspace{-0.1cm}
Our algorithm can be further enhanced along the following lines. First, the cost function of robust MPC/CCO is assumed to be deterministic, i.e., they do not contain state and motion uncertainty. A straightforward way of rectifying this would be to formulate stochastic cost as constraints using some slack variables. We are also looking at more complex applications like multi-agent navigation, reinforcement learning.

\noindent \textbf{Limitations:} Our formulations can fail due to two reasons. First, there can be a failure in the construction of the desired distribution if no feasible solution exists for the optimization problem proposed in (\ref{cost_scenario})-(\ref{feasible_scenario}). Second, if the control sampling is not dense enough, and we miss some out on some feasible control inputs.  We can handle the latter by increasing the resolution of the discretization of control inputs, although this would lead to an increase in computation time.

\noindent \textbf{Extensions to multi-agent scenario:} Our formulation can be easily extended to multiple decision-making agents. Essentially, instead of building our chance-constrained optimization and its reformulations over the VO constraint (\ref{vo}), we would construct its distributed multi-agent variant called reciprocal velocity obstacle \cite{rvo}. The construction of the desired distribution would change accordingly. The remaining steps of our GMM or RKHS based formulation remain the same.

\vspace{-0.4cm}
\appendix
\section*{Constructing the desired distribution} \label{sec:appendix-desired-distribution}
% \vspace{-0.15cm}
We now describe how distributions $P_{\textbf{w}_{t-1}}^{des}$, $P_{{_{j}^{o}}\boldsymbol{\xi}_t}^{des}$ and $P^{des}_{f_t^j}$ can be computed. While exact computations may be intractable, we provide a simple way of constructing an approximate estimate of these distributions. The basic procedure is as follows. 
Given $n$ samples of $\textbf{w}_{t-1}, {_{j}^{o}}\boldsymbol{\xi}_t$ we construct two sets  $\mathcal{C}_{\widetilde{\textbf{w}}_{t-1}}$, $\mathcal{C}_{\widetilde{{_{j}^{o}}\boldsymbol{\xi}}_t}$  containing $n_r$ samples of $\textbf{w}_{t-1}$ and $n_o$ samples of ${_{j}^{o}}\boldsymbol{\xi}_t$ respectively. For clarity of exposition, we choose $\widetilde{\textbf{w}}_{t-1}^p$, $\widetilde{{_{j}^{o}}\boldsymbol{\xi}}_t^q$ to respectively identify $p^{th}, q^{th}$ samples from set $\mathcal{C}_{\widetilde{\textbf{w}}_{t-1}}$, $\mathcal{C}_{\widetilde{{_{j}^{o}}\boldsymbol{\xi}}_t}$. Now, assume that the following holds.

\vspace{-0.35cm}
\small
\begin{equation}
f_t^j(\widetilde{\textbf{w}}_{t-1}^p, \widetilde{{_{j}^{o}}\boldsymbol{\xi}}_{t}^q, \textbf{u}_{t-1}^{nom}  )\leq 0, \forall \widetilde{\textbf{w}}_{t-1}^p   \in \mathcal{C}_{\widetilde{\textbf{w}}_{t-1}},  \widetilde{{_{j}^{o}}\boldsymbol{\xi}}_t^q\in \mathcal{C}_{\widetilde{{_{j}^{o}}\boldsymbol{\xi}}_t}
\label{desired_sample_distri}
\end{equation}  
\normalsize
\vspace{-0.45cm}

By comparing (\ref{desired_chance}) and (\ref{desired_sample_distri}) it can be inferred that the sets $\mathcal{C}_{\widetilde{\textbf{w}}_{t-1}}$, $\mathcal{C}_{\widetilde{{_{j}^{o}}\boldsymbol{\xi}}_t}$ are in fact sample approximations of the distributions $P_{\textbf{w}_{t-1}}^{des}$ and $P_{{_{j}^{o}}\boldsymbol{\xi}_{t-1}}^{des}$ respectively. Furthermore, a set $\mathcal{C}_{f_t^j}$ containing $n_r*n_o$ samples of $f_t^j(\widetilde{\textbf{w}}_{t-1}^p, \widetilde{{_{j}^{o}}\boldsymbol{\xi}}_{t}^q, \textbf{u}_{t-1}^{nom}  )$ can  be taken as the sample approximation of the desired  distribution $P^{des}_{f_t^j}$.

One major issue is which $n_r$ samples of $\textbf{w}_{t-1}$ and $n_o$ samples of ${_{j}^{o}}\boldsymbol{\xi}_t$ should be chosen to construct sets $\mathcal{C}_{\widetilde{\textbf{w}}_{t-1}}$, $\mathcal{C}_{\widetilde{{_{j}^{o}}\boldsymbol{\xi}}_t}$. In particular, we need to ensure that the chosen samples indeed satisfy the assumption (\ref{desired_sample_distri}). To this end, we follow the following process. We arbitrarily choose $n_r$ samples of $\textbf{w}_{t-1}$ and $n_o$ samples of ${_{j}^{o}}\boldsymbol{\xi}_t$ and correspondingly obtain a suitable $\textbf{u}_{t-1}^{nom}$ as a solution to the following optimization problem. 

\vspace{-0.4cm}
\small
\begin{subequations}
\begin{align}
\textbf{u}_{t-1}^{nom} = \arg\min J(\textbf{u}_{t-1})\label{cost_scenario}\\
f_t^j(\widetilde{\textbf{w}}_{t-1}^p, \widetilde{{_{j}^{o}}\boldsymbol{\xi}}_{t}^q, \textbf{u}_{t-1}^{nom}  )\leq 0, \forall p = {1,2..n_r},q = {1,2..n_o} \label{scenario_const}\\
\textbf{u}_{t-1}\in \mathcal{C}\label{feasible_scenario}
\end{align}
\end{subequations}
\normalsize
\vspace{-0.45cm}
%For example, in our implementation, we compute the left hand side of (\ref{scenario_const}) for different combination of samples and then choose the set which leads to the least violation of the constraints (\ref{scenario_const})

Note that satisfaction of (\ref{scenario_const}) ensures that the assumption (\ref{desired_sample_distri}) holds. Few points are worth noting about the above optimization. First, it is a deterministic problem whose complexity  primarily depends on the algebraic nature of $f_t^j(.)$. Second, the desired distribution  can always be constructed if we have access to sets $\mathcal{C}_{\widetilde{\textbf{w}}_{t-1}}$, $\mathcal{C}_{\widetilde{{_{j}^{o}}\boldsymbol{\xi}}_t}$. The construction of these two sets is guaranteed as long as we can obtain a feasible solution to (\ref{cost_scenario})-(\ref{feasible_scenario}). Although it is difficult to provide solution guarantees, in our simulations and experimental runs, we have observed that the solution could always be obtained if there existed a collision avoiding control considering only the mean of the uncertainty. Finally, (\ref{cost_scenario})-(\ref{feasible_scenario}) is precisely the so-called scenario approximation for the robust MPC (\ref{cost_mpc})-(\ref{feasible_mpc}). Conventionally, scenario approximation is solved with a large $n_r, n_o$ (typically $10^2$ each ) in order to obtain a solution that satisfy chance constraints (\ref{chance_mpc}) with a high $\eta$ ($\approx 0.90$). In contrast, we use (\ref{cost_scenario})-(\ref{feasible_scenario}) to estimate the desired distribution and thus for our purpose, a small sample size in the range of $n_r=n_o \approx 20$ proves to be sufficient in practice.

\ifCLASSOPTIONcaptionsoff
  \newpage
\fi

% \bibliographystyle{IEEEtran}  
% \bibliography{IEEEabrv,references} 

% \newrefsection

\onecolumn
\begin{center}
    {\Huge Supplementary Material}
\end{center}

\vspace{1cm}

The supplementary material has detailed discussions on
\begin{itemize}
    \item Derivations for expressing MMD (\ref{kernel_trick1}) in terms of kernel matrices
    \item Guarantees on safety
    \item Computational aspects of the proposed approach
    \item Additional comparisons with state of the art methods
\end{itemize}

\textbf{The videos for all the simulations and the real time runs can be found at }

\begin{center}

\href{https://robotics.iiit.ac.in/uploads/Main/Publications/rkhs-collision-avoidance/}{\textit{https://robotics.iiit.ac.in/uploads/Main/Publications/rkhs-collision-avoidance/}}
    
\end{center}

\title{Test}
\maketitle

\section*{Derivations: Expressing MMD (\ref{kernel_trick1}) in terms of Kernel matrices}

\subsection*{Prerequisites}
\subsubsection*{RKHS}
RKHS is a Hilbert space with a positive definite function $ k(\cdot) :\  \mathbb{R}^n \times \mathbb{R}^N \to \mathbb{R} $ called the Kernel. Let, $\textbf{x}$ denote an observation in physical space (say Euclidean). It is possible to embed this observation in the RKHS by defining the following kernel based function whose first argument is always fixed at $\textbf{x}$.

{\small
\begin{equation}
      \phi (\textbf{x}) = k(\textbf{x}, \cdot)
      \label{eq:rkhs-derivation-kernel}
\end{equation}}

An attractive feature of RKHS is that it is endowed with an inner product, which in turn, can be used to model the distance between two functions in RKHS. Furthermore, the distance can be formulated in terms of the kernel function in the following manner.

{\small
\begin{equation}
    \langle \phi( \textbf{x}_i ), \phi( \textbf{x}_j ) \rangle = k(\textbf{x}_i, \textbf{x}_j)
    \label{eq:rkhs-derivation-kernel-inner-product}
\end{equation}}

The equation (\ref{eq:rkhs-derivation-kernel-inner-product}) is called the "kernel trick" and its strength lies in the fact that the inner product can be computed by only pointwise evaluation of the kernel function.

\vspace{0.4cm}
\subsubsection*{Distribution Embedding of PVO}
Let $ \textbf{w}^1_{t-1}, \textbf{w}^2_{t-1}, \ldots, \textbf{w}^n_{t-1}$ be samples drawn from a distribution corresponding to the uncertainty of the robot's state and control. Similarly, let $ {_j^o}\boldsymbol{\xi}_{t}^1, {_j^o}\boldsymbol{\xi}_{t}^2, \ldots, {_j^o}\boldsymbol{\xi}_{t}^n $ be samples drawn from the distribution corresponding to the uncertainty in the state of the obstacle. It is important to note here that the parametric forms of these distributions need not be known. Both these distributions (of the robot and the obstacle) can be represented in the RKHS through a function called the kernel mean, which is described as

{\small
\begin{subequations}
\begin{align}
     \label{eq:kernel-derivation-kme-definition}
     \mu[\textbf{w}_{t-1}] = \sum^n_{p=1} \alpha_p k(\textbf{w}^p_{t-1}, \cdot) \\
      \mu[{_j^o}\boldsymbol{\xi}_{t}] = \sum^n_{q=1} \beta_q k({_j^o}\boldsymbol{\xi}_{t}^q, \cdot)
\end{align}
\end{subequations}}

where $ \alpha_p $ is the weight associated with $ \textbf{w}^p_{t-1} $ and $ \beta_q $ is the weight associated with $ {_j^o}\boldsymbol{\xi}_{t}^q $. For example, if the samples are $i.i.d$, then $ \alpha_p = \frac{1}{n} \forall p $.

Following [refer], equation (\ref{eq:kernel-derivation-kme-definition}) can be used to embed functions of random variables like $ f^j_t(\textbf{w}_{t-1}, {_j^o}\boldsymbol{\xi}_{t}, \textbf{u}_{t-1}) $ shown in equation (\ref{eq:kernel-derivation-kme}), where $ f_t^j $ is the VO constraint and $ \textbf{w}_{t-1}, {_j^o}\boldsymbol{\xi}_{t} $ are random variables with definitions taken from table \ref{symbols}.

{\small
\begin{equation}
     \label{eq:kernel-derivation-kme}
     \mu_{f^j_t}(\textbf{u}_{t-1}) = \sum^n_{p=1} \sum^n_{q=1} \alpha_p \beta_q k(f^j_t (\textbf{w}^p_{t-1}, {_j^o}\boldsymbol{\xi}_{t}^q, \textbf{u}_{t-1}), \cdot)
\end{equation}}

Similarly, we can write the same for the samples for the desired distribution,

{\small
\begin{equation}
     \sum^{n_r}_{p=1} \sum^{n_o}_{q=1} \lambda_p \psi_q k(f^j_t (\widetilde{\textbf{w}^p}_{t-1}, \widetilde{{_j^o}\boldsymbol{\xi}_{t}^q}, \textbf{u}^{nom}_{t-1}), \cdot)
\end{equation}}

where $  \alpha_p, \beta_q, \lambda_p $ and $ \psi_q $ are the constants obtained from a reduced method that is explained in the section \ref{reduced_set}.

An important point to notice from above is that for given samples $ \textbf{w}_{t-1}, {_j^o}\boldsymbol{\xi}_{t} $ the kernel mean is dependent on variable $ \textbf{u}_{t-1} $. The rest of the material focuses on the detailed derivation of the $ \mathcal{L}_{dist} $ used in our cost function to solve our robust MPC problem.

The proposed method has its foundations built by first posing the robust MPC as a moment matching problem (Theorem \ref{th_moment_bound}) and then describes a solution that is a workaround based on the concept of embedding distributions in RKHS and Maximum Mean Discrepancy (MMD). Further insights on how MMD can act as a surrogate to the moment matching problem are described in Theorem \ref{th_poly_mmd} of the paper. Theorem \ref{th_poly_mmd} suggests that if two distributions $ P_{f^j_t} $ and $ P^{des}_{f^j_t} $, have their distributions embedded in RKHS for a polynomial kernel up to order $d$, then decreasing the MMD distance becomes a way to match the first $d$ moments of these distributions.

Using these insights, an optimization problem given by (\ref{cost_mpc}), (\ref{cost_reform_gen})-(\ref{feas_reform_gen}) is proposed.

{\small
\begin{subequations}
\begin{align}
    \arg \min \rho_1 \mathcal{L}_{dist} + \rho_2 J(\textbf{u}_{t-1}) \\
    \mathcal{L}_{dist} = \Vert \mu_{{P}_{f^j_t}}(\textbf{u}_{t-1}) - \mu_{{P}^{des}_{f^j_t}} \Vert^2 \\
    J(\textbf{u}_{t-1}) = \Vert\boldsymbol{\overline{\xi}}_{t} -   \boldsymbol{\xi}_t^d\Vert_2^2 +\Vert \textbf{u}_{t-1}\Vert_2^2 \\
    \textbf{u}_{t-1} \in \mathcal{C}
\end{align}
\end{subequations}}

where,

{\small
\begin{equation}
\begin{split}
\label{eq:kernel-derivations-mmd-definition}
    \Vert \mu_{{P}_{f^j_t}}(\textbf{u}_{t-1}) - \mu_{{P}^{des}_{f^j_t}} \Vert^2 & =
    \langle \mu_{{P}_{f^j_t}}(\textbf{u}_{t-1}), \mu_{{P}_{f^j_t}}(\textbf{u}_{t-1}) \rangle -
    2 \langle \mu_{{P}_{f^j_t}}(\textbf{u}_{t-1}), \mu_{{P}^{des}_{f^j_t}} \rangle +
    \langle \mu_{{P}^{des}_{f^j_t}}, \mu_{{P}^{des}_{f^j_t}} \rangle \\
    & = \langle \sum^n_{p=1} \sum^n_{q=1} \alpha_p \beta_q k(f^j_t (\textbf{w}^p_{t-1},
        {_j^o}\boldsymbol{\xi}_{t}^q, \textbf{u}_{t-1}), \cdot), \sum^n_{p=1} \sum^n_{q=1}
        \alpha_p \beta_q k(f^j_t (\textbf{w}^p_{t-1}, {_j^o}\boldsymbol{\xi}_{t}^q,
        \textbf{u}_{t-1}), \cdot) \rangle \\
        & - 2 \langle \sum^n_{p=1} \sum^n_{q=1} \alpha_p \beta_q k(f^j_t (\textbf{w}^p_{t-1},
        {_j^o}\boldsymbol{\xi}_{t}^q, \textbf{u}_{t-1}), \cdot), \sum^{n_r}_{p=1} \sum^{n_o}_{q=1}
        \lambda_p \psi_q k(f^j_t (\widetilde{\textbf{w}^p}_{t-1}, \widetilde{{_j^o}\boldsymbol{\xi}_{t}^q},
        \textbf{u}^{nom}_{t-1}), \cdot) \rangle \\
        & + \langle \sum^{n_r}_{p=1} \sum^{n_o}_{q=1} \lambda_p \psi_q k(f^j_t
            (\widetilde{\textbf{w}^p}_{t-1}, \widetilde{{_j^o}\boldsymbol{\xi}_{t}^q},
            \textbf{u}^{nom}_{t-1}), \cdot),
            \sum^{n_r}_{p=1} \sum^{n_o}_{q=1} \lambda_p \psi_q k(f^j_t
            (\widetilde{\textbf{w}^p}_{t-1}, \widetilde{{_j^o}\boldsymbol{\xi}_{t}^q},
            \textbf{u}^{nom}_{t-1}), \cdot) \rangle
\end{split}
\end{equation}}

Using kernel trick, the expression in (\ref{eq:kernel-derivations-mmd-definition}) can be reduced to

{\small
\begin{equation}
\begin{split}
    \Vert \mu_{{P}_{f^j_t}}(\textbf{u}_{t-1}) - \mu_{{P}^{des}_{f^j_t}} \Vert^2 & = \
    \textbf{C}_{\alpha \beta} \textbf{K}_{f^j_t f^j_t} \textbf{C}_{\alpha \beta}^T \
    - \ 2 \textbf{C}_{\alpha \beta} \textbf{K}_{f^j_t \widetilde{f^j_t}} \textbf{C}_{\lambda \psi}^T \
    + \ \textbf{C}_{\lambda \psi} \textbf{K}_{\widetilde{f^j_t} \widetilde{f^j_t}} \textbf{C}_{\lambda \psi}^T
\end{split}
\end{equation}}

where,

{\small
\begin{equation}
\begin{split}
    \textbf{C}_{\alpha \beta} & = \ \begin{bmatrix}
        \alpha_1 \beta_1 & \alpha_2 \beta_2 & \cdots & \alpha_n \beta_n
    \end{bmatrix}_{1 \times n^2} \\
    \textbf{C}_{\lambda \psi} & = \ \begin{bmatrix}
        \lambda_1 \psi_1 & \lambda_2 \psi_2 & \cdots & \lambda_n \psi_n
        \end{bmatrix}_{1 \times n_r n_o} \\
    \textbf{K}_{f^j_t f^j_t} & = \ \begin{bmatrix}
        \textbf{K}^{11}_{f^j_t f^j_t} & \textbf{K}^{12}_{f^j_t f^j_t} & \cdots & \textbf{K}^{1n}_{f^j_t f^j_t} \\
        \textbf{K}^{21}_{f^j_t f^j_t} & \textbf{K}^{22}_{f^j_t f^j_t} & \cdots & \textbf{K}^{2n}_{f^j_t f^j_t} \\
        \vdots & \vdots & \ddots & \vdots \\
        \textbf{K}^{n1}_{f^j_t f^j_t} & \textbf{K}^{n2}_{f^j_t f^j_t} & \cdots & \textbf{K}^{nn}_{f^j_t f^j_t} \\
    \end{bmatrix} \\
    \textbf{K}^{ab}_{f^j_t f^j_t} & = \ \begin{bmatrix}
        k(f^j_t (\textbf{w}^a_{t-1}, {_j^o}\boldsymbol{\xi}_{t}^1, \textbf{u}_{t-1}),
        f^j_t (\textbf{w}^b_{t-1}, {_j^o}\boldsymbol{\xi}_{t}^1, \textbf{u}_{t-1})) & \cdots &
        k(f^j_t (\textbf{w}^a_{t-1}, {_j^o}\boldsymbol{\xi}_{t}^1, \textbf{u}_{t-1}),
        f^j_t (\textbf{w}^b_{t-1}, {_j^o}\boldsymbol{\xi}_{t}^n, \textbf{u}_{t-1})) \\
        \vdots & \ddots & \vdots \\
        k(f^j_t (\textbf{w}^a_{t-1}, {_j^o}\boldsymbol{\xi}_{t}^n, \textbf{u}_{t-1}),
        f^j_t (\textbf{w}^b_{t-1}, {_j^o}\boldsymbol{\xi}_{t}^1, \textbf{u}_{t-1})) & \cdots &
        k(f^j_t (\textbf{w}^a_{t-1}, {_j^o}\boldsymbol{\xi}_{t}^n, \textbf{u}_{t-1}),
        f^j_t (\textbf{w}^b_{t-1}, {_j^o}\boldsymbol{\xi}_{t}^n, \textbf{u}_{t-1}))
    \end{bmatrix}_{n \times n} \\
    \textbf{K}_{f^j_t \widetilde{f^j_t}} & = \ \begin{bmatrix}
        \textbf{K}^{11}_{f^j_t \widetilde{f^j_t}} & \textbf{K}^{12}_{f^j_t \widetilde{f^j_t}} & \cdots & \textbf{K}^{1n_r}_{f^j_t \widetilde{f^j_t}} \\
        \textbf{K}^{21}_{f^j_t \widetilde{f^j_t}} & \textbf{K}^{22}_{f^j_t \widetilde{f^j_t}} & \cdots & \textbf{K}^{2n_r}_{f^j_t \widetilde{f^j_t}} \\
        \vdots & \vdots & \ddots & \vdots \\
        \textbf{K}^{n1}_{f^j_t \widetilde{f^j_t}} & \textbf{K}^{n2}_{f^j_t \widetilde{f^j_t}} & \cdots & \textbf{K}^{nn_r}_{f^j_t \widetilde{f^j_t}} \\
    \end{bmatrix} \\
    \textbf{K}^{ab}_{f^j_t \widetilde{f^j_t}} & = \ \begin{bmatrix}
        k(f^j_t (\textbf{w}^a_{t-1}, {_j^o}\boldsymbol{\xi}_{t}^1, \textbf{u}_{t-1}),
        f^j_t (\widetilde{\textbf{w}^b}_{t-1}, \widetilde{{_j^o}\boldsymbol{\xi}_{t}^1}, \textbf{u}^{nom}_{t-1}))
        & \cdots &
        k(f^j_t (\textbf{w}^a_{t-1}, {_j^o}\boldsymbol{\xi}_{t}^1, \textbf{u}_{t-1}),
        f^j_t (\widetilde{\textbf{w}^b}_{t-1}, \widetilde{{_j^o}\boldsymbol{\xi}_{t}^{n_o}}, \textbf{u}^{nom}_{t-1})) \\
        \vdots & \ddots & \vdots \\
        k(f^j_t (\textbf{w}^a_{t-1}, {_j^o}\boldsymbol{\xi}_{t}^n, \textbf{u}_{t-1}),
        f^j_t (\widetilde{\textbf{w}^b}_{t-1}, \widetilde{{_j^o}\boldsymbol{\xi}_{t}^1}, \textbf{u}^{nom}_{t-1}))
        & \cdots &
        k(f^j_t (\textbf{w}^a_{t-1}, {_j^o}\boldsymbol{\xi}_{t}^n, \textbf{u}_{t-1}),
        f^j_t (\widetilde{\textbf{w}^b}_{t-1}, \widetilde{{_j^o}\boldsymbol{\xi}_{t}^{n_o}}, \textbf{u}^{nom}_{t-1}))
    \end{bmatrix}_{n \times n_o}
\end{split}
\end{equation}}
\section*{Guarantees on Safety}
Both our GMM-KLD and RKHS based approach works with only sample level information without assuming any parametric form for the underlying distribution. Thus, the performance guarantees on safety depends on the following aspects. First, on how well are we modeling the distribution of our collision avoidance function (PVO) for a given finite sample size. Second, does our modeling improve as the samples increase: a property popularly known as consistency in estimation. Third, can we tune our model to produce diverse trajectories with varied probability of avoidance in line with the original robust MPC formulation. The discussion regarding the consistency, has been addressed in section \ref{sec:performance-guarantees}. The third question has already been addressed in Remark \ref{gmm_rkhs_tuning} and additional results validating this are provided in the following subsections. Moreover, first two questions about GMM-KLD based approaches have already been established in the existing literature \cite{GMM-fit-quality},\cite{GMM-fit-quality1}. 

\subsection*{Trajectory Tuning}
\begin{figure*}[h]
    \centering
    \subfigure[]{
        \centering
        \includegraphics[width=0.3\textwidth]{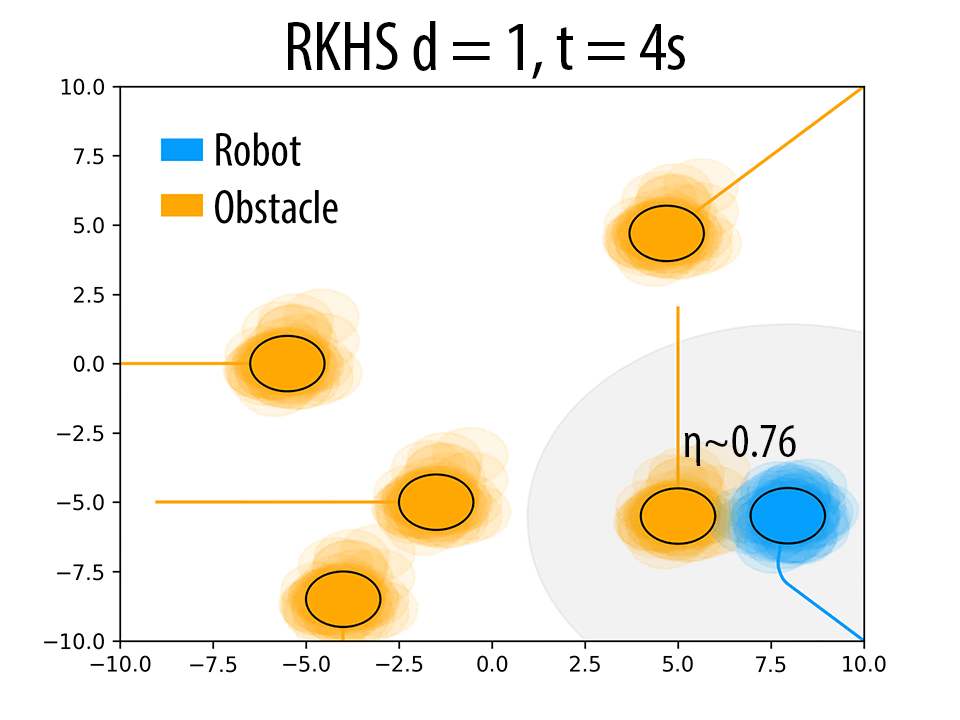}
    }
    \subfigure[]{
        \centering
        \includegraphics[width=0.3\textwidth]{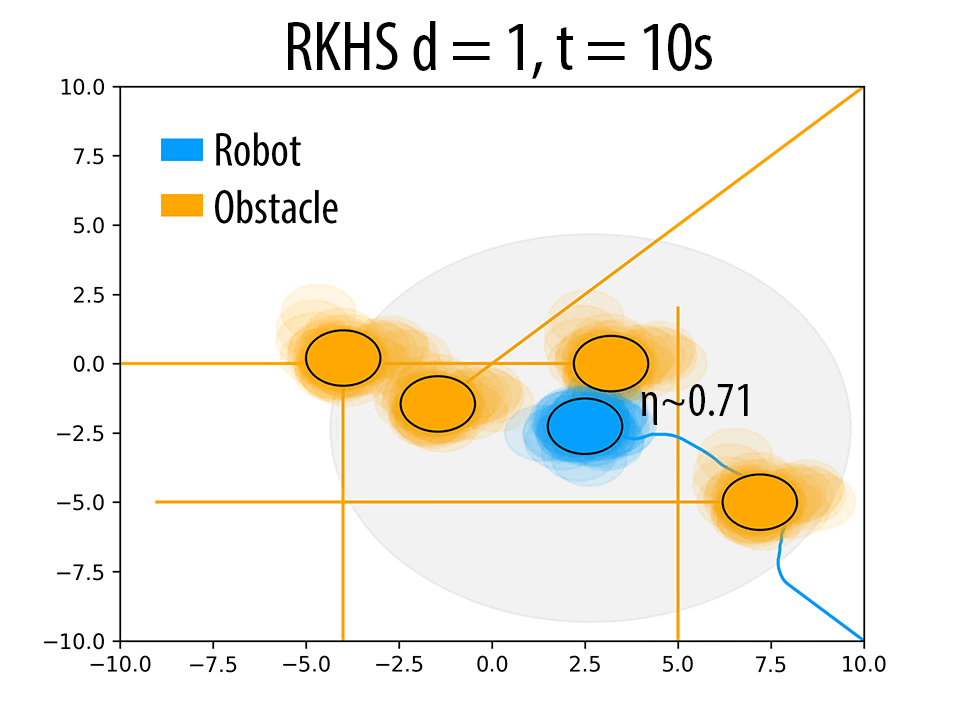}
    }
    \subfigure[]{
        \centering
        \includegraphics[width=0.3\textwidth]{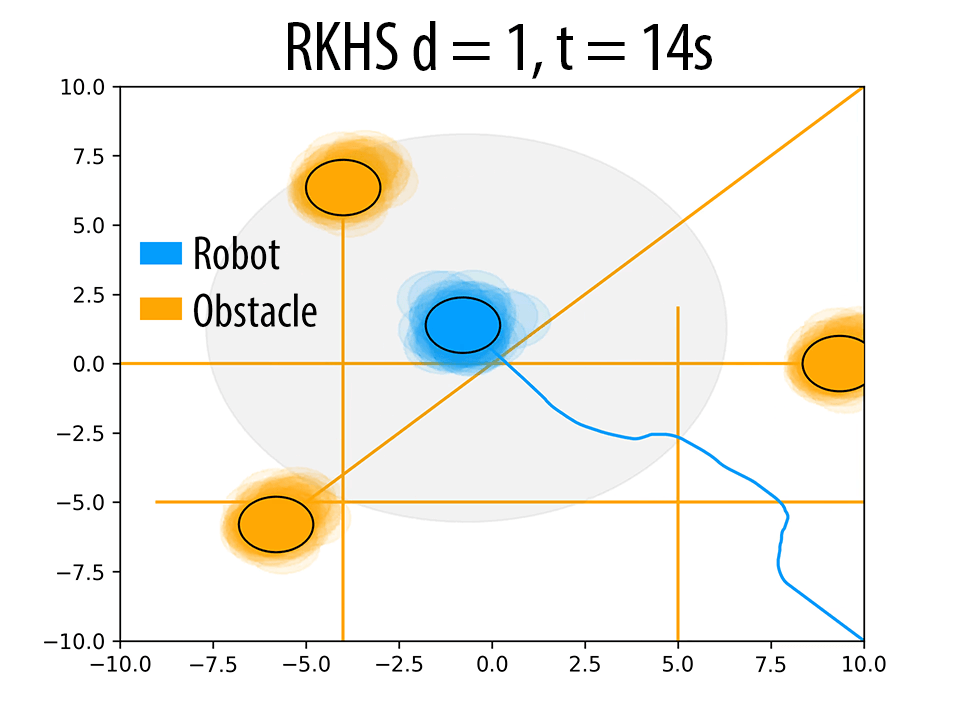}
    }
    \subfigure[]{
        \centering
        \includegraphics[width=0.3\textwidth]{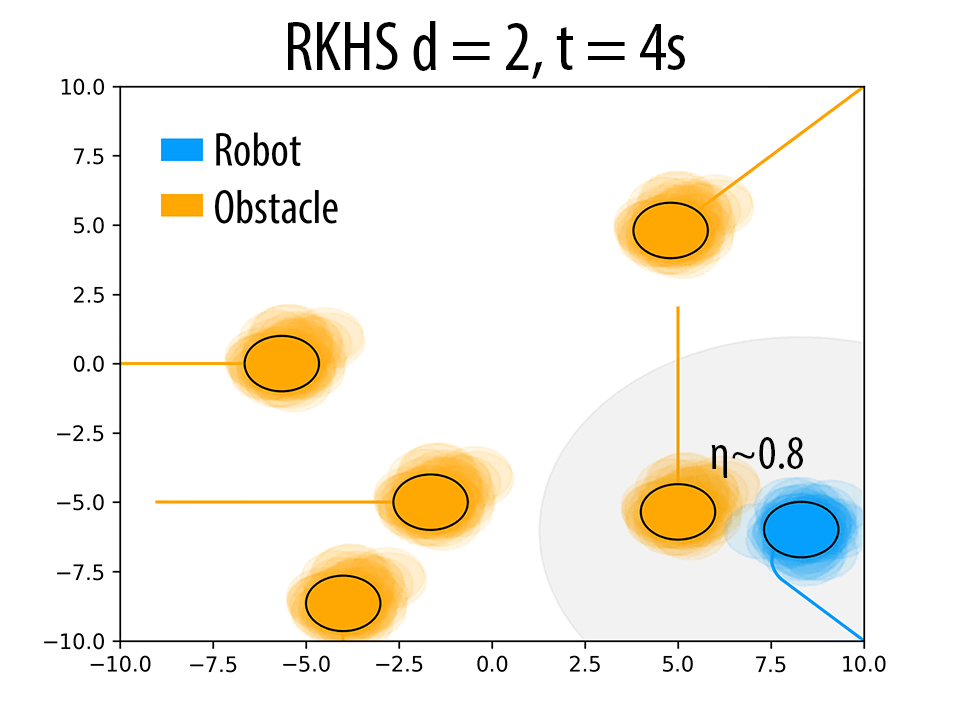}
    }
    \subfigure[]{
        \centering
        \includegraphics[width=0.3\textwidth]{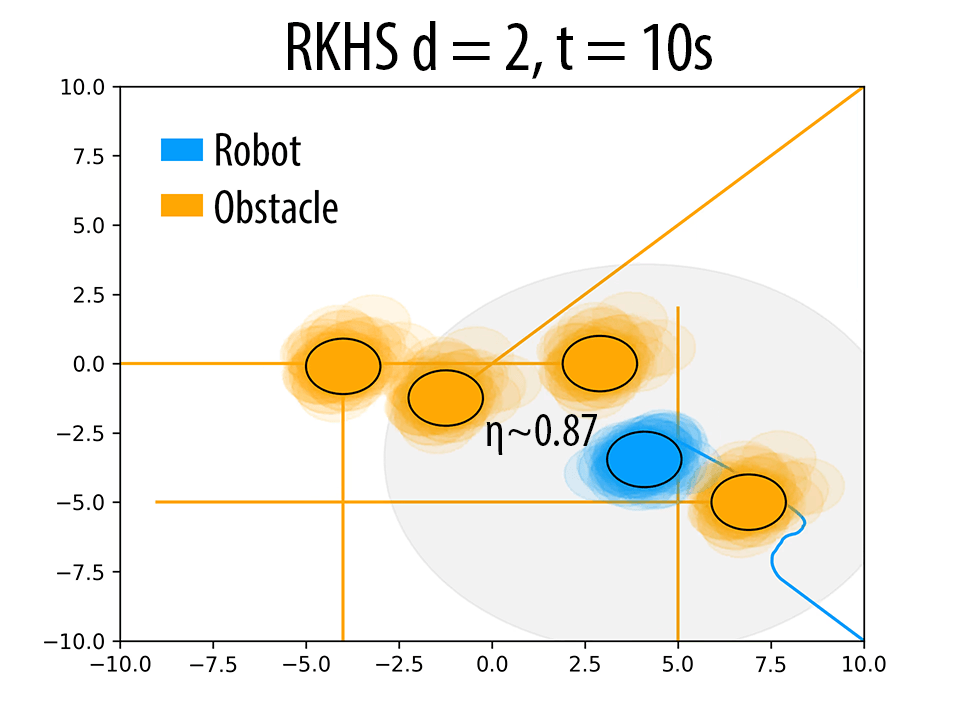}
    }
    \subfigure[]{
        \centering
        \includegraphics[width=0.3\textwidth]{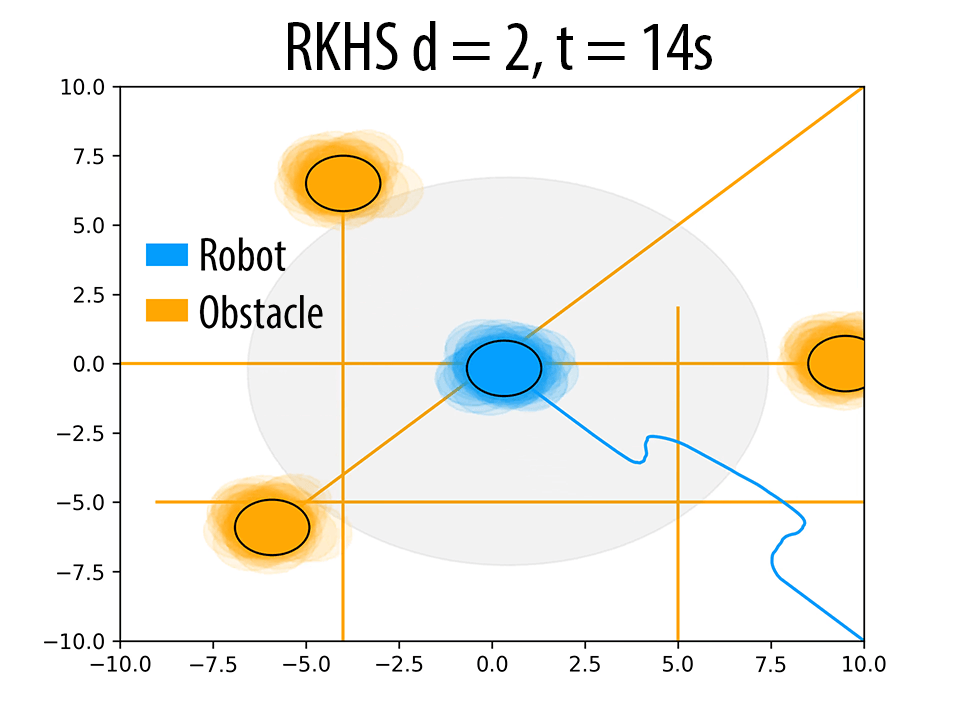}
    }
    \subfigure[]{
        \centering
        \includegraphics[width=0.3\textwidth]{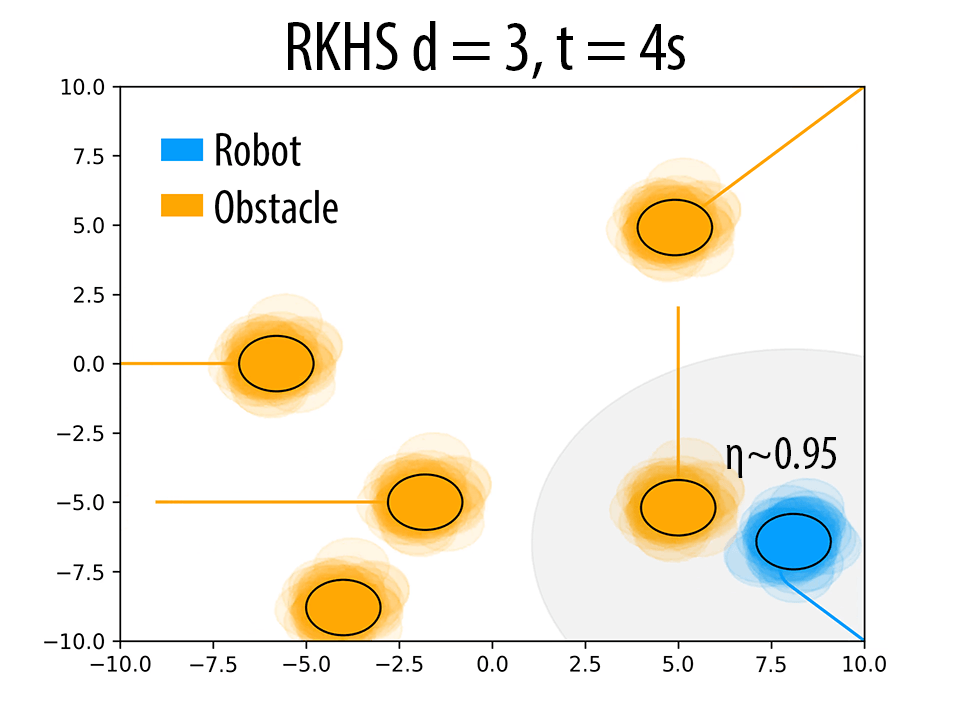}
    }
    \subfigure[]{
        \centering
        \includegraphics[width=0.3\textwidth]{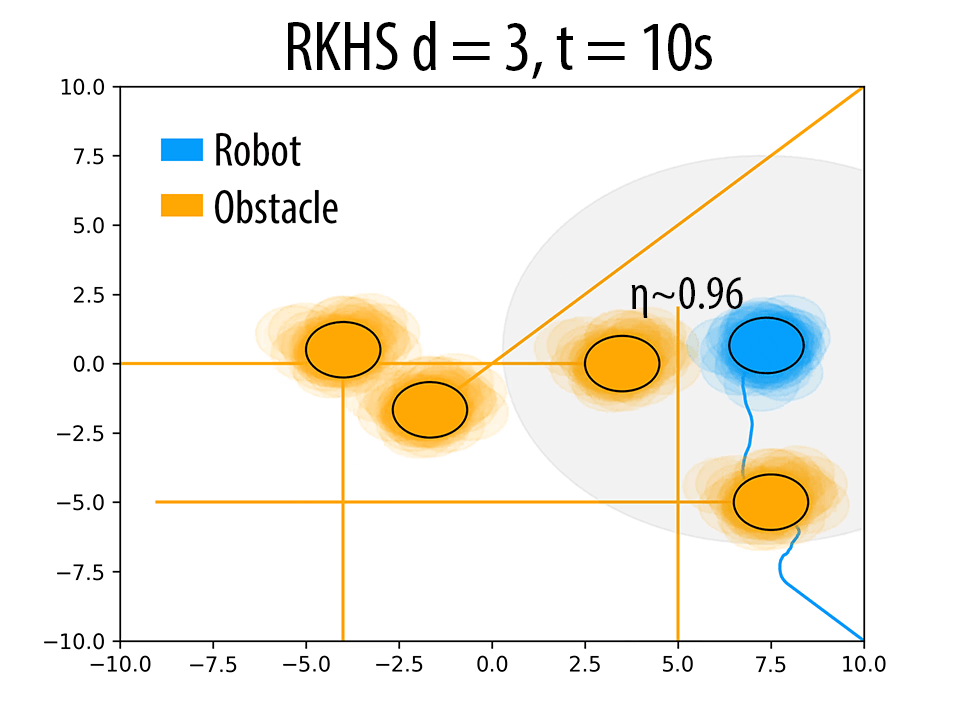}
    }
    \subfigure[]{
        \centering
        \includegraphics[width=0.3\textwidth]{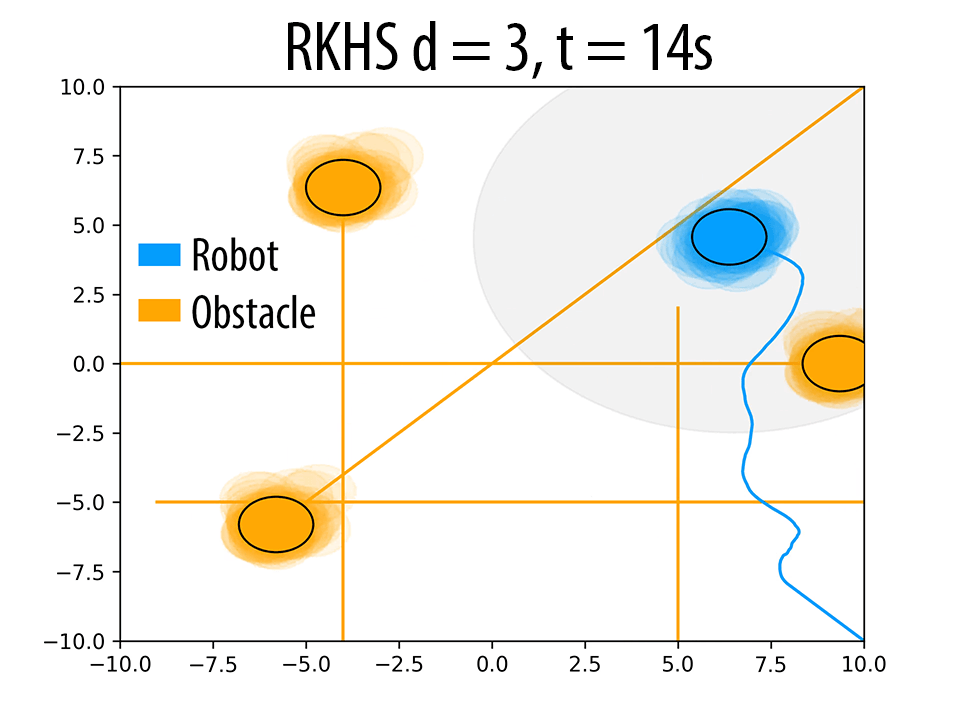}
    }
    \caption{Effect of $d$ on the trajectory of the robot. $\eta$ increases with increase in $d$. In figures (a),(d) and (g) (at t = 4s), for a similar configuration, with increasing d, the number of samples of robot and obstacle that overlap decreases, thereby resulting in higher $\eta$ values. Similar trend is observed in b, e and h (at t = 10s). The different trajectories resulted due to different d values are shown in figures (c),(f) and (i) (at t = 14s).}
    \label{fig:rebuttal-rkhs-multi-obs-1}
\end{figure*}

As mentioned in Remark \ref{gmm_rkhs_tuning} and also validated in the results Section \ref{sec:simulation-validation}, the polynomial kernel order $d$ can be used to produce trajectories with different probability of collision avoidance $\eta$. Herein, we presents some additional results to back our claims. Fig.\ref{fig:rebuttal-rkhs-multi-obs-1} summarizes these results. As $d$ increases, the robot seeks to maintain higher clearance from the obstacles.

\subsection*{Analysis of real runs}
\begin{figure*}[h]
    \centering
    \subfigure[]{
        \centering
        \includegraphics[width=0.23\textwidth]{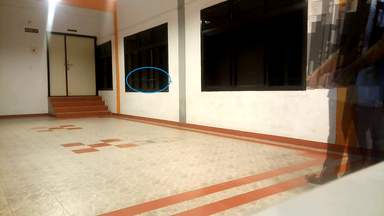}
    }
    \subfigure[]{
        \centering
        \includegraphics[width=0.23\textwidth]{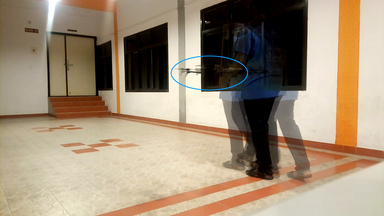}
    }
    \subfigure[]{
        \centering
        \includegraphics[width=0.23\textwidth]{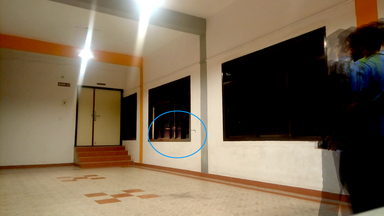}
    }
    \subfigure[]{
        \centering
        \includegraphics[width=0.23\textwidth]{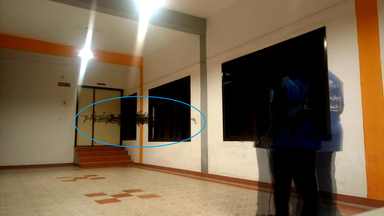}
    }
    \caption{The ghost figures of the person represent various samples of the moving person being considered while avoiding. Similarly, the ghost figures of the drone is indicative of the various samples of the drone being considered while performing the maneuver. In (b) and (d) the drone starts avoiding the person. The deviation while executing the avoidance maneuver is higher in case of d = 3. Further analysis is presented in the Figs. \ref{fig:rebuttal-real-runs-trajectory} and \ref{fig:rebuttal-real-runs-dists}.}
    \label{fig:rebuttal-real-runs-snaps}
\end{figure*}

The RKHS framework was implemented on a Bebop drone equipped with a monocular camera. A person walking with an April tag marker in front of him constitute the moving object. Distance and bearing to the marker is computed using the April Tag library from which the velocity information is obtained. The state and velocity of the drone is obtained from the on board odometry (estimated using an onboard IMU). The state, velocity, control and perception/measurement noise distributions were computed through a non-parametric Pearson fit over experimental data obtained over multiple runs. A number of experimental runs, totally 15, were performed to evaluate the RKHS method. This is shown in the Fig. \ref{fig:rebuttal-real-runs-snaps}. We show that increasing the power of the polynomial kernel, d, decreases the overlap of the samples of the robot and the obstacle leading to more conservative maneuvers. 

\begin{figure*}[h]
    \centering
    \subfigure[]{
        \centering
        \includegraphics[width=0.45\textwidth]{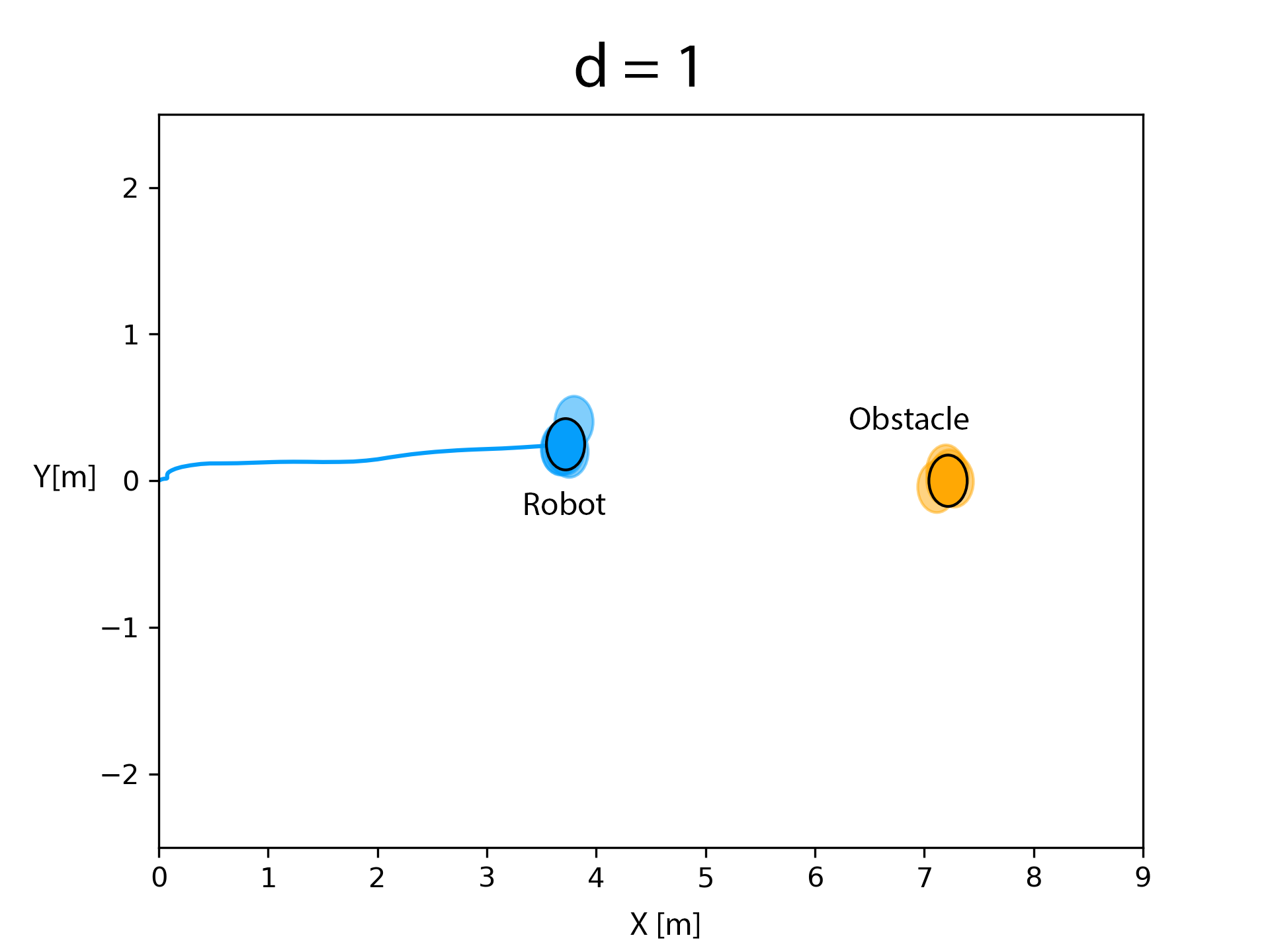}
    }
    \subfigure[]{
        \centering
        \includegraphics[width=0.45\textwidth]{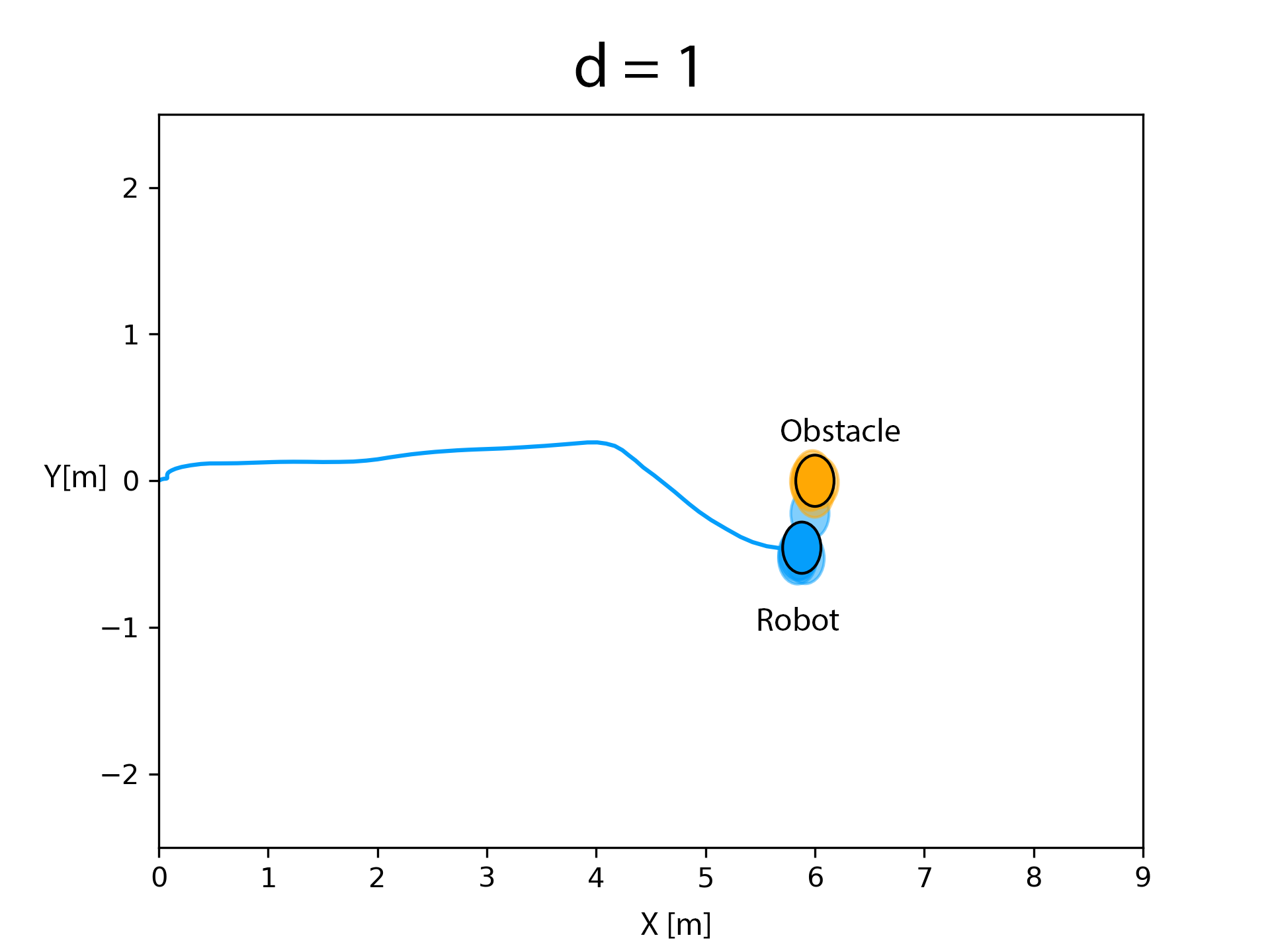}
    }
    \subfigure[]{
        \centering
        \includegraphics[width=0.45\textwidth]{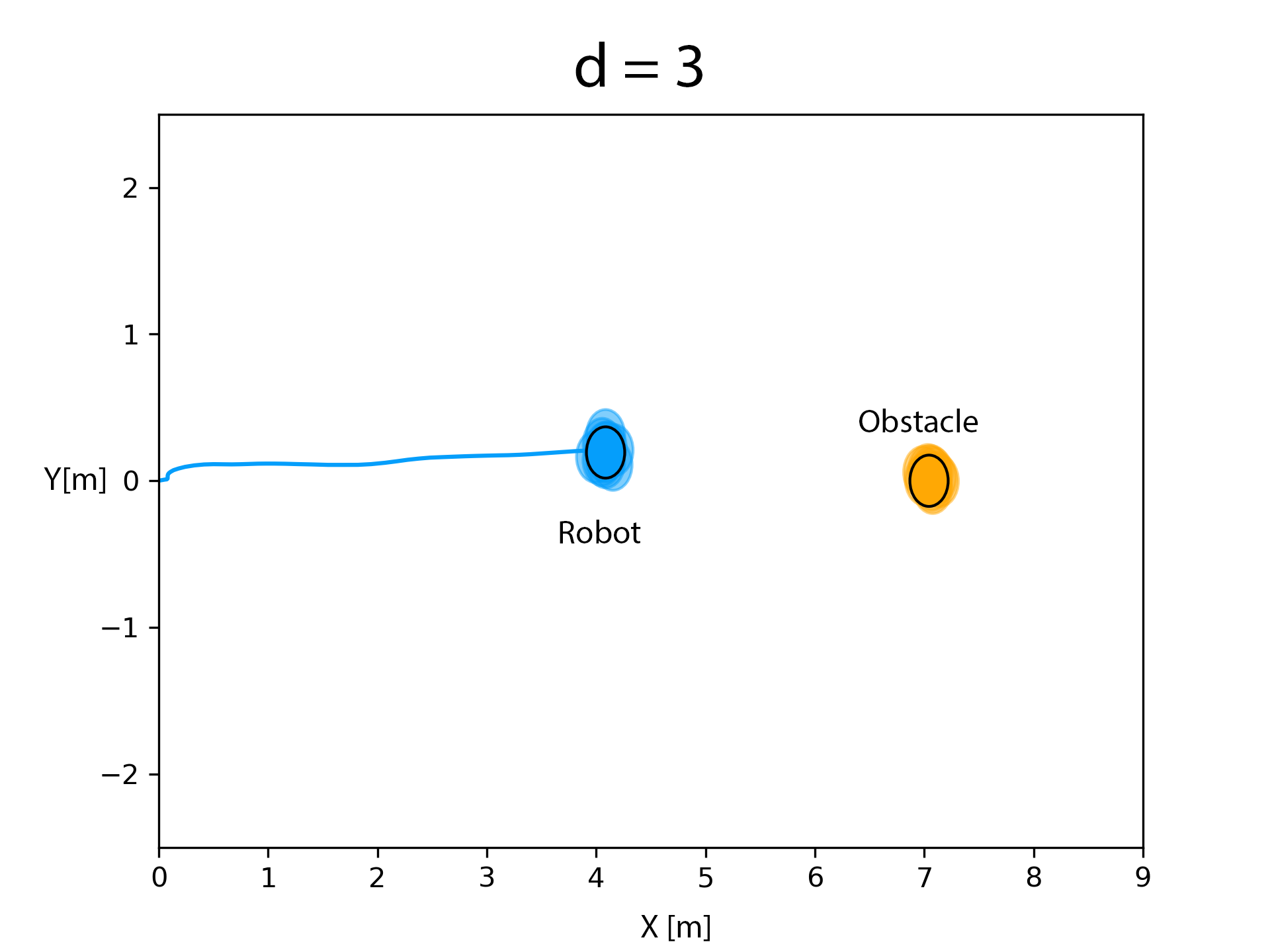}
    }
    \subfigure[]{
        \centering
        \includegraphics[width=0.45\textwidth]{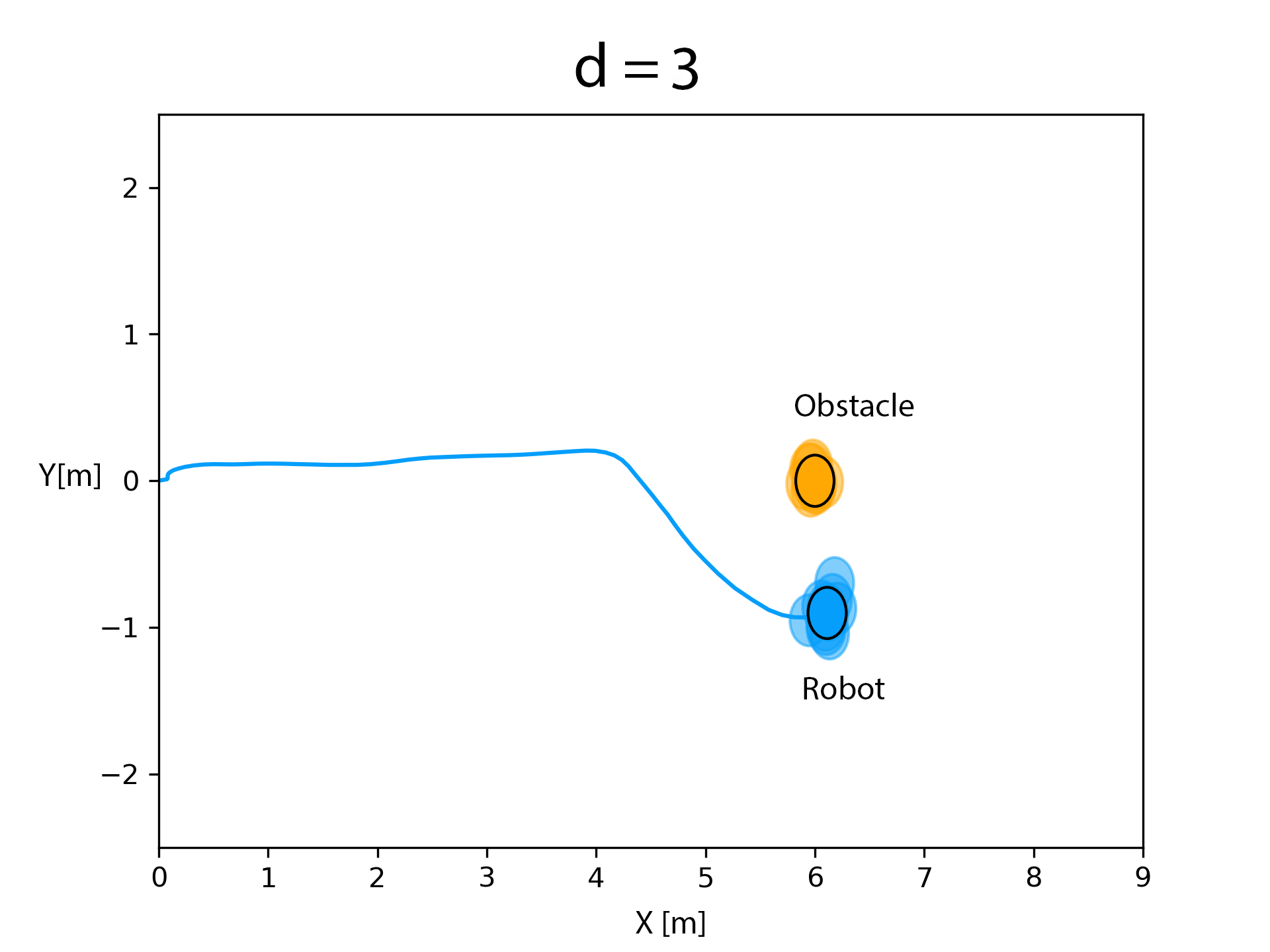}
    }
    \caption{Trajectory plots resulting from real time collision avoidance shown in Fig. \ref{fig:rebuttal-real-runs-snaps} (a)-(b) show the trajectories obtained for $d=1$. (c)-(d) show trajectories obtained with $d=3$. As can be seen, higher $d$ reduces the overlap between robot and obstacle uncertainty leading to higher probability of collision avoidance.}
    \label{fig:rebuttal-real-runs-trajectory}
\end{figure*}

\begin{figure*}[h]
    \centering
    \subfigure[]{
        \centering
        \includegraphics[width=0.45\textwidth]{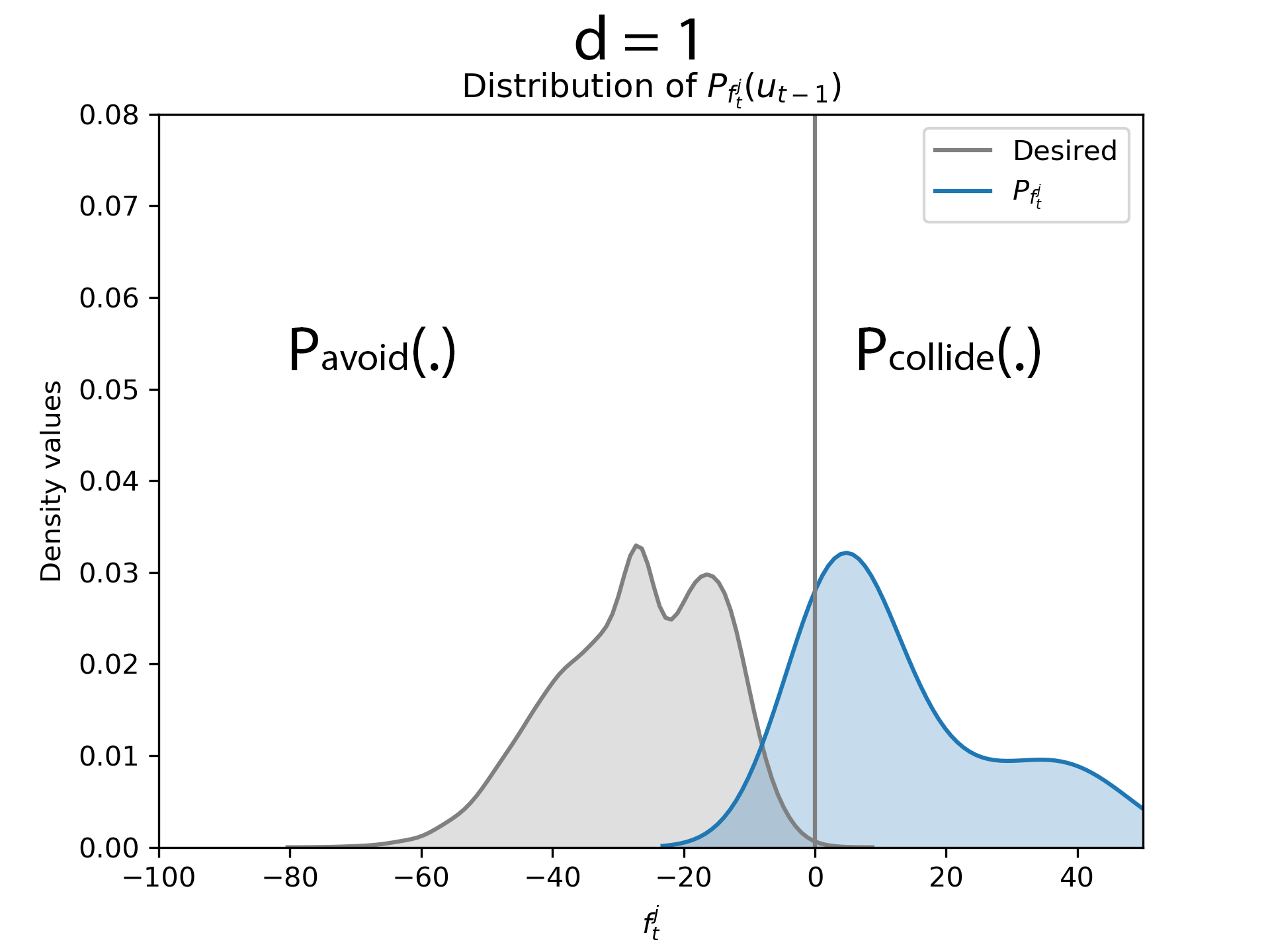}
    }
    \subfigure[]{
        \centering
        \includegraphics[width=0.45\textwidth]{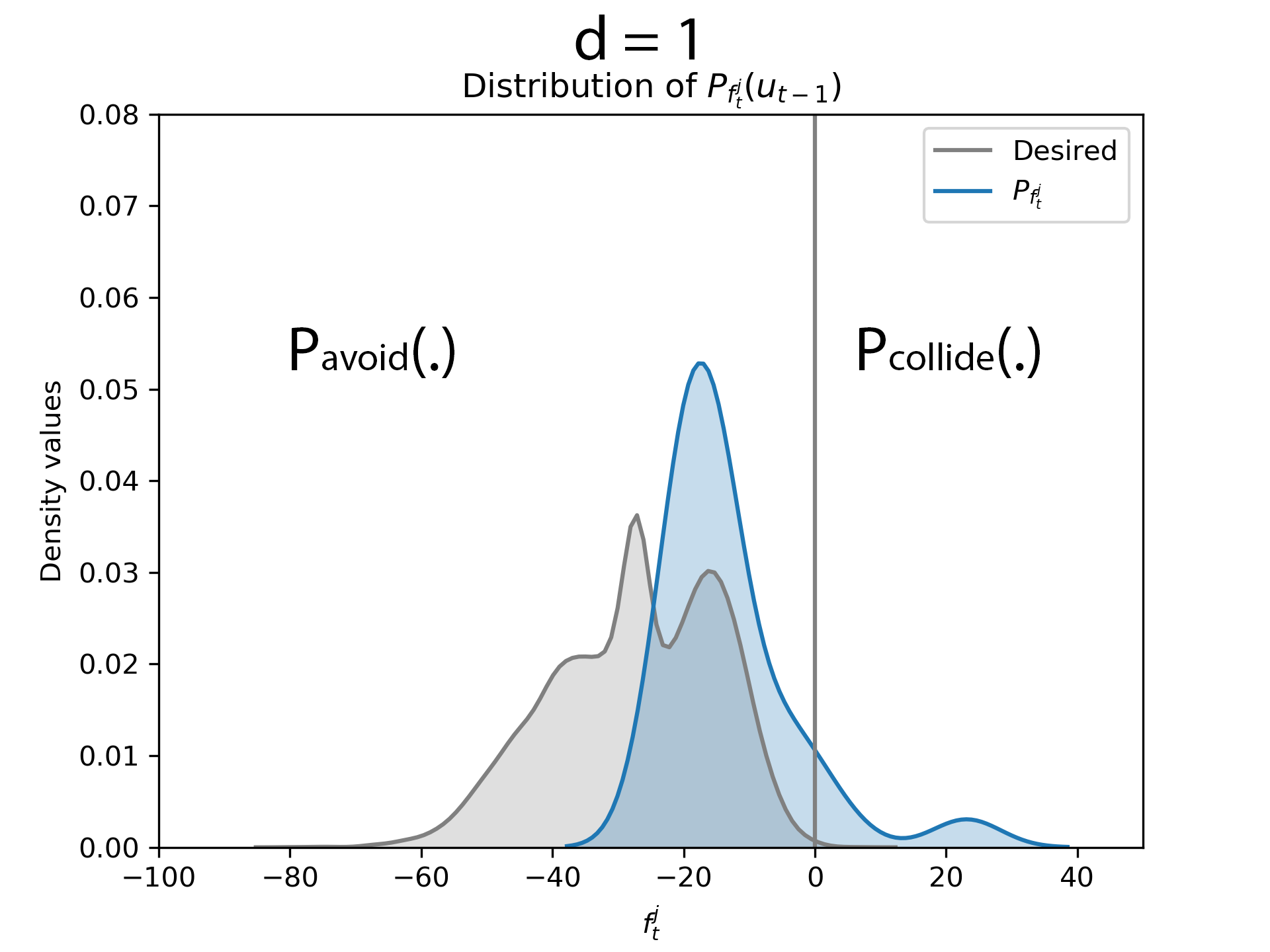}
    }
    \caption{Effect of $d$ on the distribution matching for the real runs shown in the Fig. \ref{fig:rebuttal-real-runs-snaps}. Figure (a) illustrates the behaviour of  $ P_{f^j_t} $  when the proposed algorithm is evaluated for $d=1$ while (b) shows the results for $d=3$. It can be observed that the controls computed for $d=1$ results in a notable amount of area of $ P_{f^j_t} $ on the right side indicating the presence of collision (can be verified in Fig \ref{fig:rebuttal-real-runs-trajectory}(b)) where as choosing $d=3$ results in almost negligible amount of area on the right side, leading to almost no colliding samples (can be verified in Fig \ref{fig:rebuttal-real-runs-trajectory}(d))
    }
    \label{fig:rebuttal-real-runs-dists}
\end{figure*}

The figures \ref{fig:rebuttal-real-runs-trajectory} and \ref{fig:rebuttal-real-runs-dists} further analyze the results of the real run. The robot samples are represented using translucent blue circles while the obstacle samples are represented using translucent yellow circles. The means of the robot and the obstacle are represented using opaque blue and yellow circles, respectively. 

\section*{Computational Aspects}
\begin{itemize}
    \item The proposed RKHS based approach avoids the need to first fit a parametric form to the distribution of PVO and the desired distribution and then compute collision avoidance control. For example, in the paper (Section III-D and section IV), we highlighted this aspect by fitting the popular Gaussian Mixture Model to both these distributions. Our RKHS based approach directly performs distribution level reasoning from the given samples.
    \item Furthermore, we exploit the kernel trick to reduce the MMD computation to just matrix multiplication which can be efficiently parallelized on GPUs (Section IV-B). The details of such implementations are given in the supplementary material [\href{https://robotics.iiit.ac.in/uploads/Main/Publications/rkhs-collision-avoidance/#implementation}{robotics.iiit.ac.in/uploads/Main/Publications/rkhs-collision-avoidance/\#implementation}].
\end{itemize}
\section*{Comparisons with the existing methods in the literature}
\subsection*{Linearized constraints with Gaussian approximation of uncertainty}

\begin{figure*}[h]
    \centering
    \includegraphics[width=\textwidth]{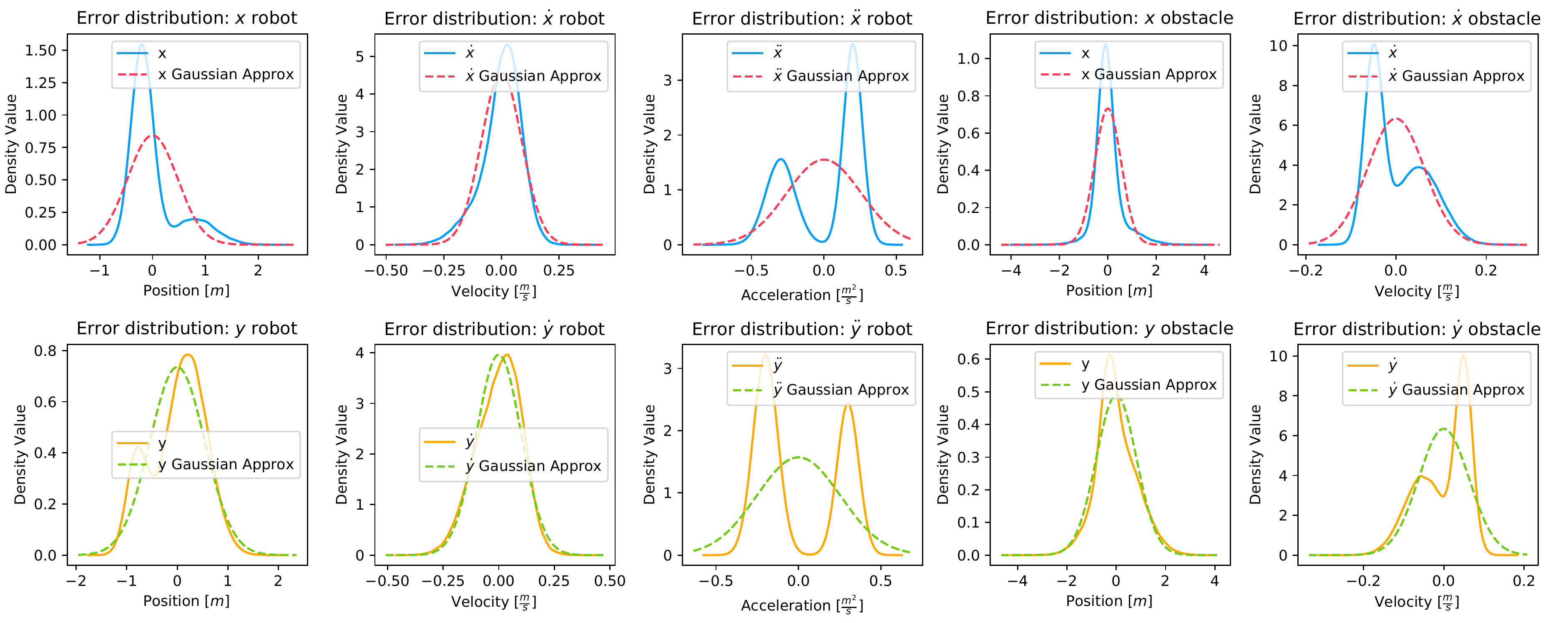}
    \caption{This figure shows the Gaussian approximations of the non parametric error distributions for position, velocity, acceleration of the robot and position, velocity of the obstacle that are shown in Fig. \ref{fig:error-distributions}. The Gaussian approximations of the non parametric distribution are shown as dotted curves.}
    \label{fig:rebuttal-noise-models}
\end{figure*}

\begin{figure*}[h]
     \centering
     \subfigure[]{
        \centering
        \includegraphics[width=0.45\textwidth]{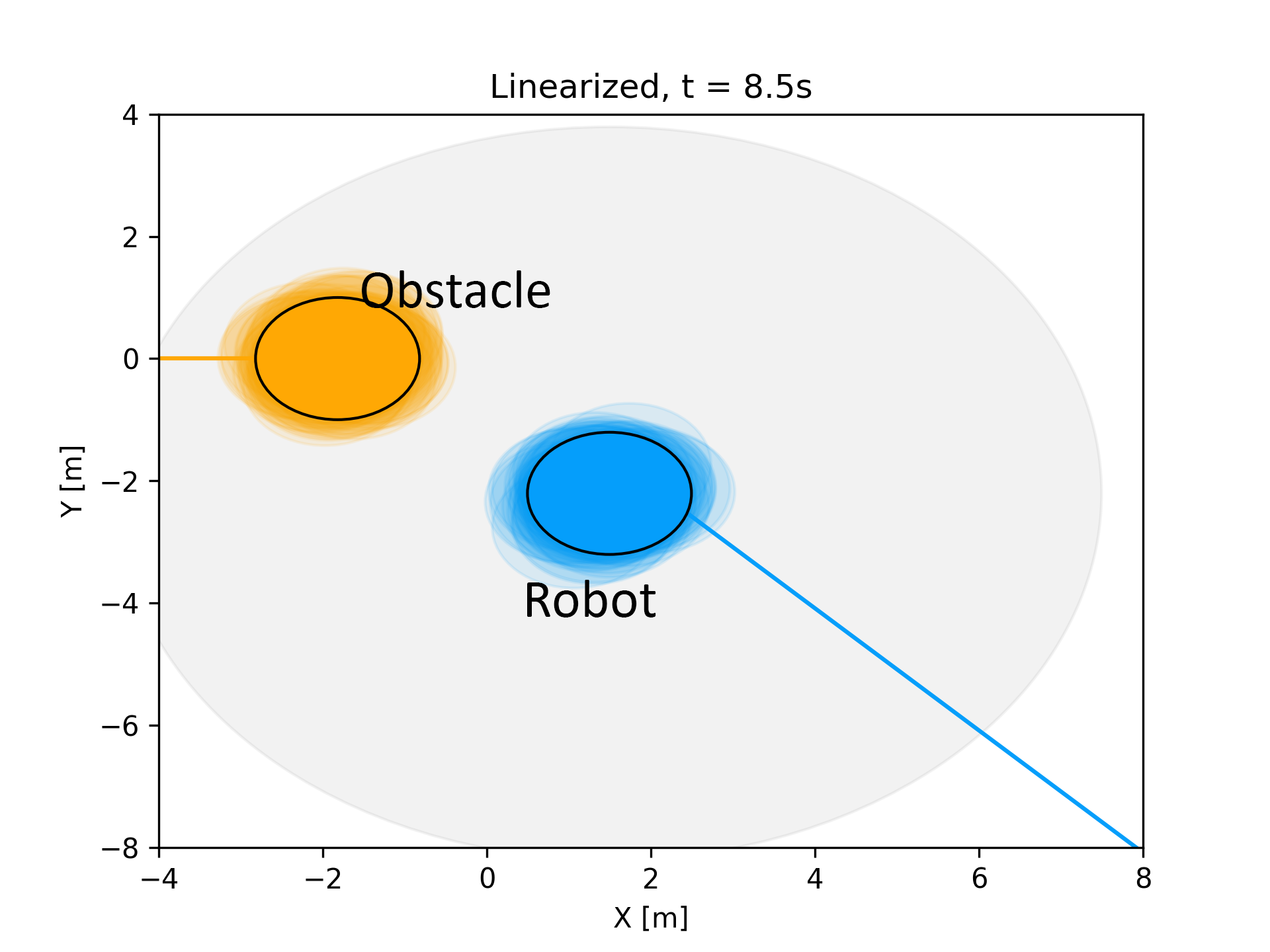}
        \label{rebuttal-lin_config}
    }
    \subfigure[]{
        \centering
        \includegraphics[width=0.45\textwidth]{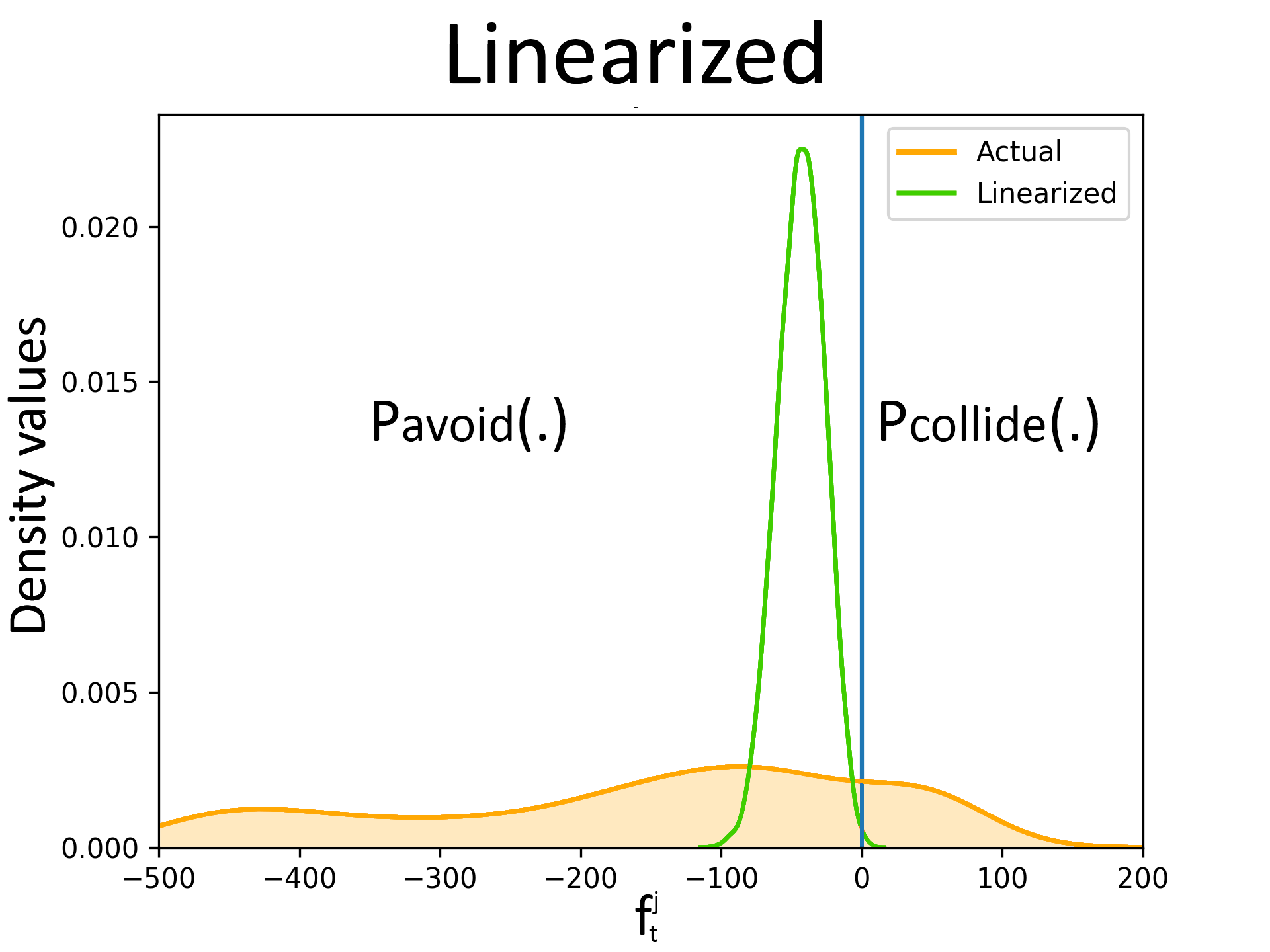}
        \label{rebuttal-lin_pdf}
    }
    \caption{We show the results for linearization+Gaussian approximation approach for the one obstacle benchmark in Fig. \ref{fig:kld-vs-gmm-one-obstacle-results}. The figure on the left shows the distribution plot for $P_{\hat{f}_t^j}(\textbf{u}_{t-1})$ while the figure on the right shows the robot (blue), obstacle (yellow), sensing range (gray). At t = 8.5s, a considerable portion of the true distribution of $P_{f^j_t}$ is on the right of zero (which means high probability of collision) while the Gaussian approximation $P_{\hat{f}^j_t}$ is completely on the left side of the zero line. This implies that $P_{\hat{f}^j_t}$ is a misrepresentation of $P_{f^j_t}$. The same can be inferred from the figure where the robot seems to be on a collision course with the obstacle. A comparison with our RKHS based approach can be found in the figure \ref{fig:rebuttal-mmd-2-traj-dist}.}
    \label{fig:rebuttal-lin-k1.0}
\end{figure*}

\begin{figure*}[h]
     \centering
     \subfigure[]{
        \centering
        \includegraphics[width=0.45\textwidth]{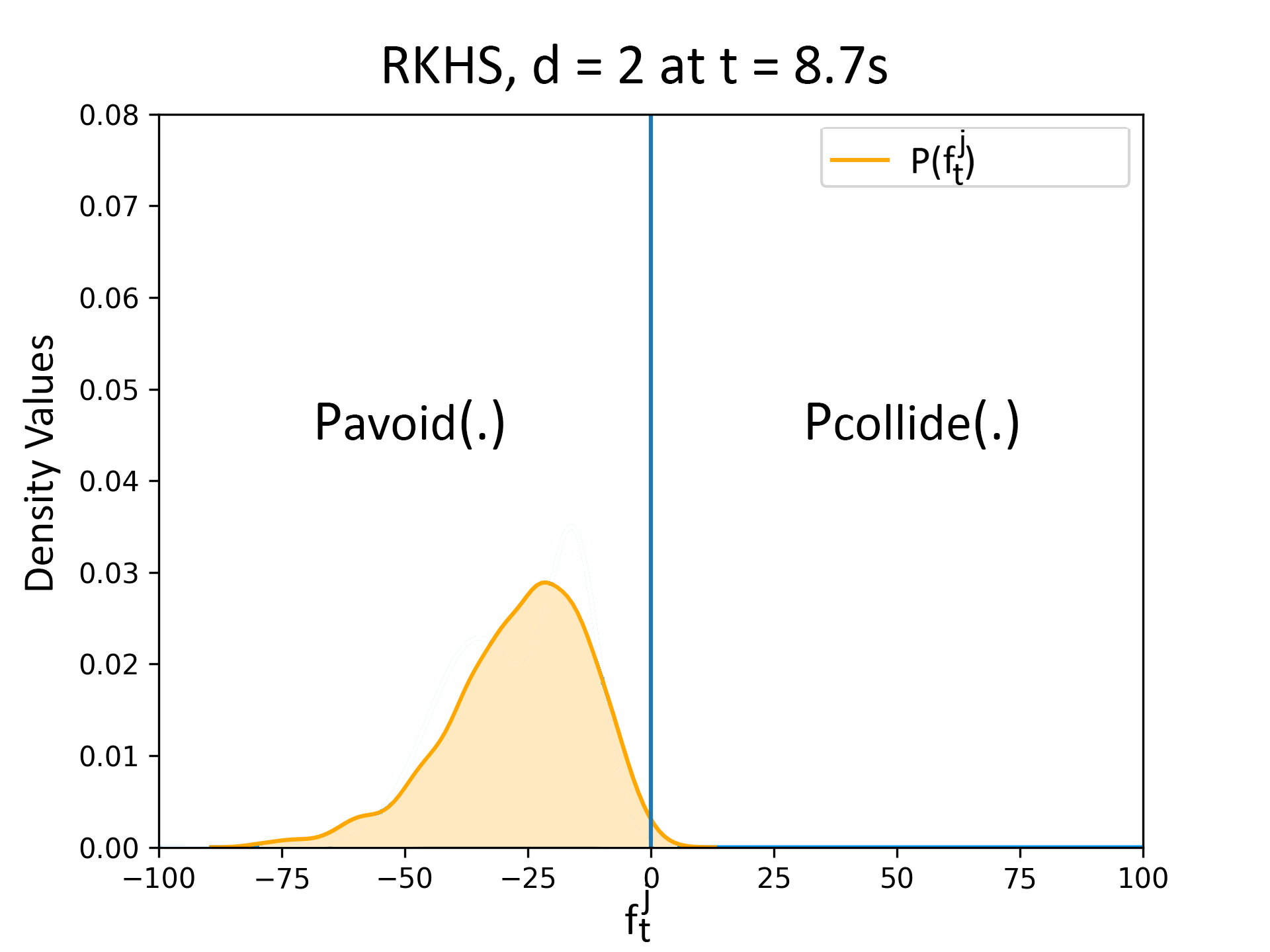}
        \label{rebuttal-rkhs_0087_dist}
    }
    \subfigure[]{
        \centering
        \includegraphics[width=0.45\textwidth]{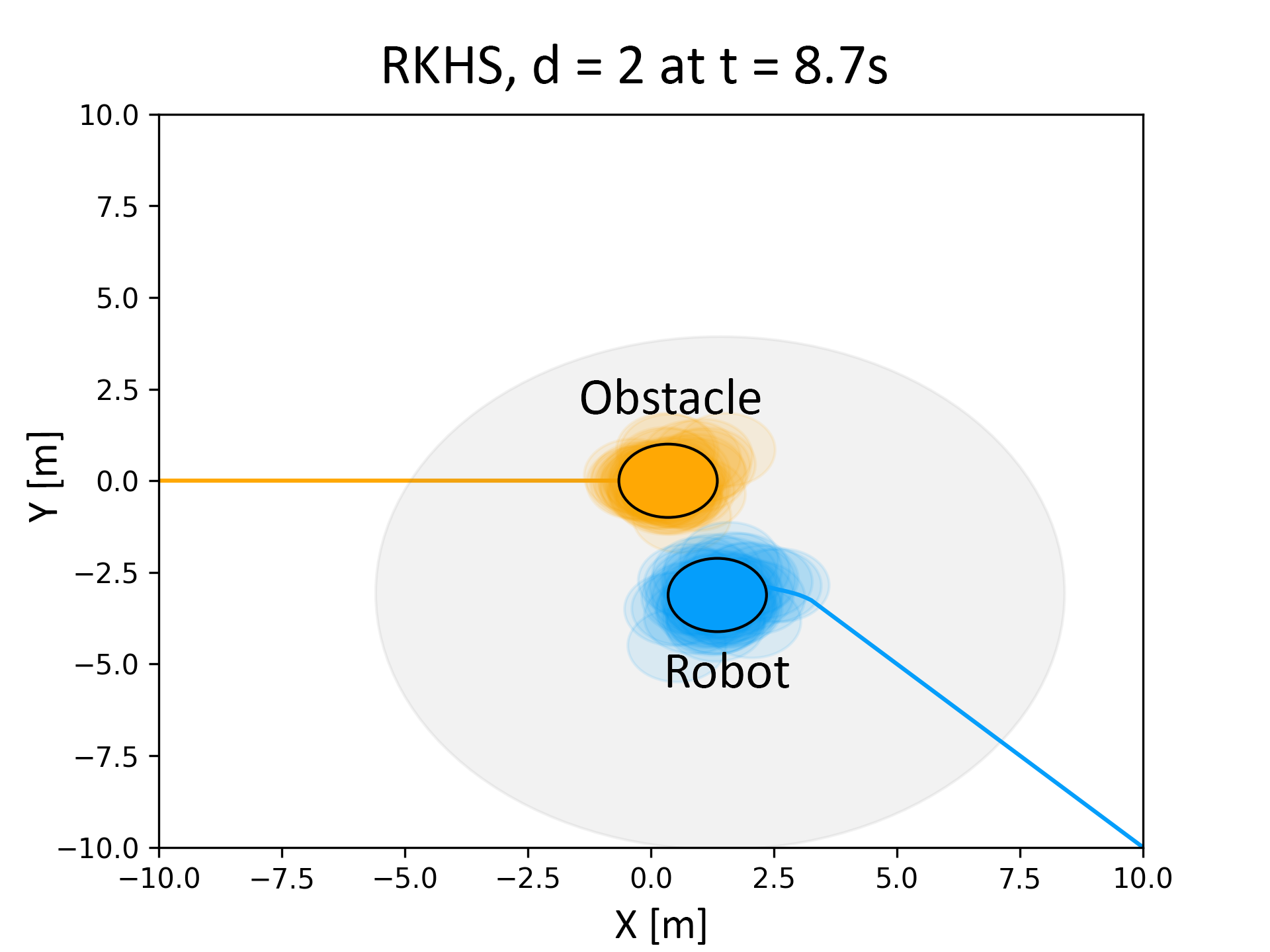}
        \label{rebuttal-rkhs_0087_config}
    }
    \caption{The figure on the left shows the distribution plots for $P_{f^j_t}$ while the figure on the right shows the robot (blue), obstacle (yellow), sensing range (gray). Unlike the case of the linearized constraints, the robot starts avoiding the obstacle in the RKHS based approach.}
    \label{fig:rebuttal-mmd-2-traj-dist}
\end{figure*}

For this comparison, we first fit a Gaussian to the error distribution (shown in Fig. \ref{fig:rebuttal-noise-models}) and consequently, approximate  $\textbf{w}_{t-1},\  {_{j}^{o}}\boldsymbol{\xi}_{t}$ as Gaussian random variables. We subsequently compute an affine approximation of $f_t^j(.)$ which we denote as  $\hat{f}_t^j(.)$ by linearizing with respect to   $\textbf{w}_{t-1}$ and ${_{j}^{o}}\boldsymbol{\xi}_{t}$. Note that the linearization is with respect to random variables and not $\textbf{u}_{t-1}$. $\hat{f}_t^j(.)$ is still non-linear in $\textbf{u}_{t-1}$. But nevertheless, $P_{\hat{f}_t^j}(\textbf{u}_{t-1})$ takes a Gaussian form \cite{alonso-mora-ral19}. We can now easily adopt sampling based procedure to obtain the control $\textbf{u}_{t-1}$ such that $P({\hat{f}^j_t(.)}\leq0)\geq\eta$ is satisfied, where $\eta$ is the probability of collision avoidance. Fig. \ref{fig:rebuttal-lin-k1.0} summarizes the results. Fig.\ref{rebuttal-lin_config} shows a robot in imminent collision with a dynamic obstacle. The true distribution $P_{f_t^j}(\textbf{u}_{t-1})$ and the Gaussian approximation $P_{\hat{f}_t^j}(\textbf{u}_{t-1})$ for the computed control input is shown in Fig.\ref{rebuttal-lin_pdf}. As shown, the Gaussian approximation lies completely to the left of $f_t^j=0$, line indicating collision avoidance with a very high probability for the computed control input. However, in reality true distribution has a significant mass to the right of  $f_t^j=0$ indicating a risk of collision. \textbf{Thus, this experiment clearly shows that the Gaussian approximation results in a collision avoidance algorithm that is not robust.}. Fig. \ref{fig:rebuttal-mmd-2-traj-dist} summarizes the collision avoidance results obtained with our RKHS based approach. As clearly shown, we can compute a control input which brings the true distribution  $P_{f_t^j}(\textbf{u}_{t-1})$ to the left of $f_t^j=0$.

\subsection*{Comparison with Gaussian approximation + exact surrogate based approach}
In this approach, we first fit a Gaussian to the error distribution (shown in the Fig. \ref{fig:rebuttal-noise-models}) and consequently, approximate  $\textbf{w}_{t-1}, {_{j}^{o}}\boldsymbol{\xi}_{t}$ as Gaussian random variables. Subsequently, we replace the chance constraints with the following surrogate used in works like \cite{musigma1},  \cite{musigma2}, \cite{chance_multi}.

\small
\begin{equation}
E[f_t^j(.)]+\epsilon\sqrt{Var[f_t^j(.)]}\leq 0, \eta\geq \frac{\epsilon^2}{1+\epsilon^2},
\label{rebuttal-meanvar}
\end{equation}
\normalsize

\noindent where, $E[f_t^j(.)]$, $Var[f_t^j(.)]$ respectively represent the expectation and variance of $f_t^j(.)$ taken with respect to random variables $\textbf{w}_{t-1}, {_{j}^{o}}\boldsymbol{\xi}_{t}$. It can be shown that the satisfaction of (\ref{rebuttal-meanvar}) ensures the satisfaction of chance constraints with confidence $\eta\geq \frac{\epsilon^2}{1+\epsilon^2}$. Note that for Gaussian uncertainty, $E[f_t^j(.)]$, $Var[f_t^j(.)]$ have a closed form expression. We henceforth call this approach EV-Gauss.

\begin{figure*}[h]
     \centering
     \subfigure{
        \centering
        \includegraphics[width=0.8\textwidth]{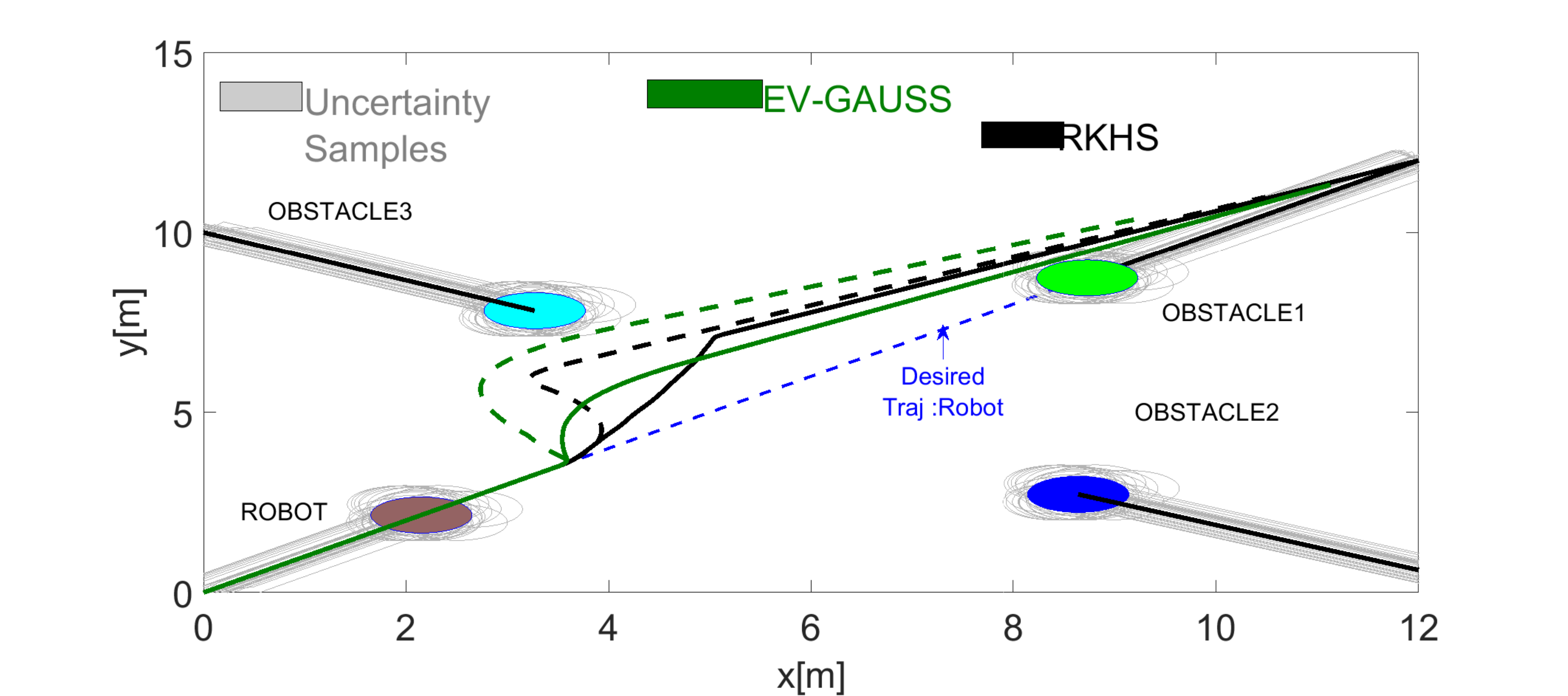}
    }
    \caption{Trajectory comparison obtained with our proposed RKHS based approach and EV Gauss. The solid lines represent trajectories for $\eta=0.60$ and the dotted lines correspond to $\eta=0.80$.  As shown, EV-Gauss approach leads to higher deviation from the desired trajectory. It can be further correlated with the results presented in Fig.\ref{fig:rebuttal-vs-evgauss-costs}}
    \label{fig:rebuttal-vs-evgauss}
\end{figure*}

\begin{figure*}[h]
    \centering
    \subfigure[]{
        \centering
        \includegraphics[width=0.45\textwidth]{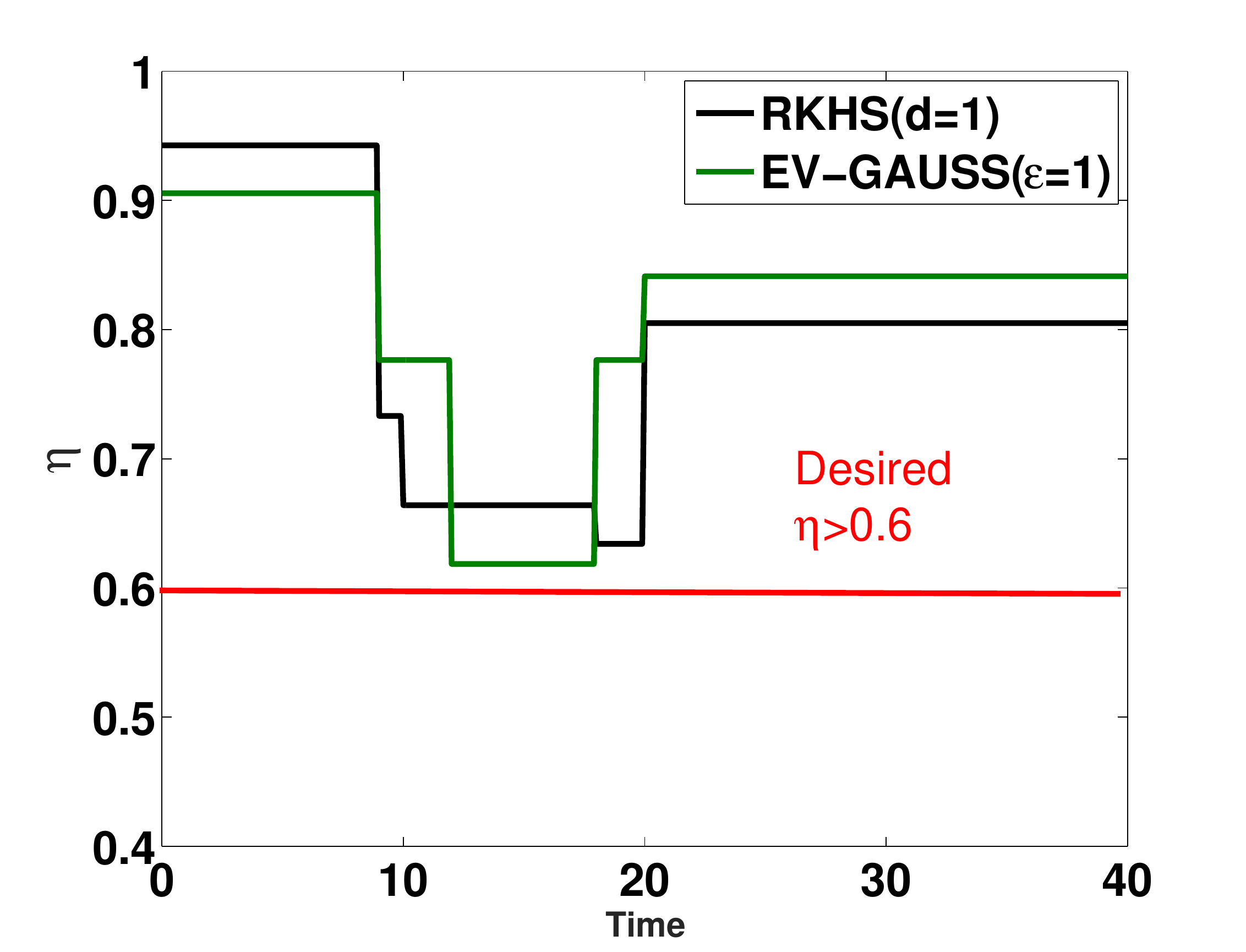}
    }
    \subfigure[]{
        \centering
        \includegraphics[width=0.45\textwidth]{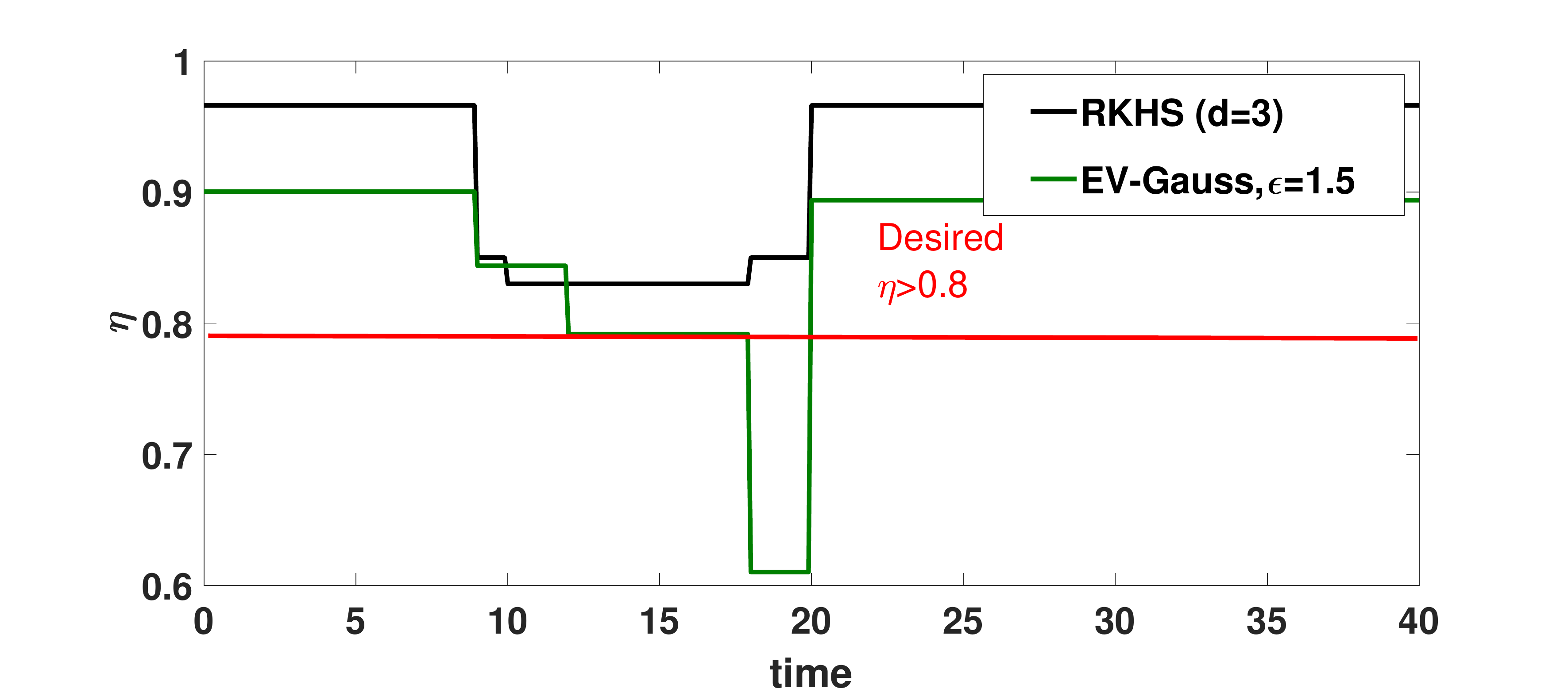}
    }
    \caption{Confidence plots for EV Gauss based approach. At high $\eta$ (Fig (b)) the EV Gauss method fails to maintain the minimum confidence required, while RKHS based method performs more reliably.}
    \label{fig:rebuttal-vs-evgauss-confidence}
\end{figure*}

\begin{figure*}[h]
    \centering
    \subfigure[]{
        \centering
        \includegraphics[width=0.3\textwidth]{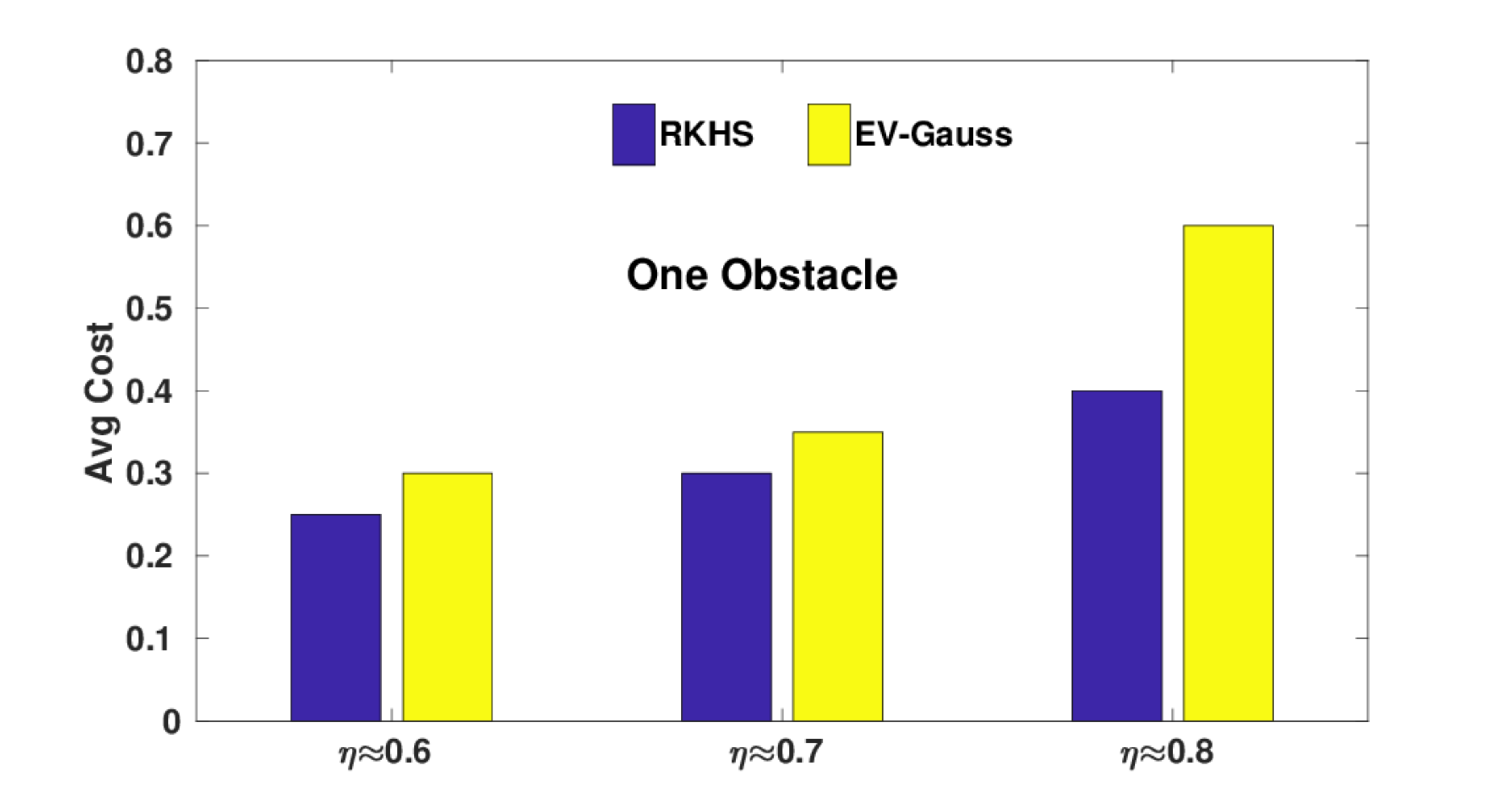}
    }
    \subfigure[]{
        \centering
        \includegraphics[width=0.3\textwidth]{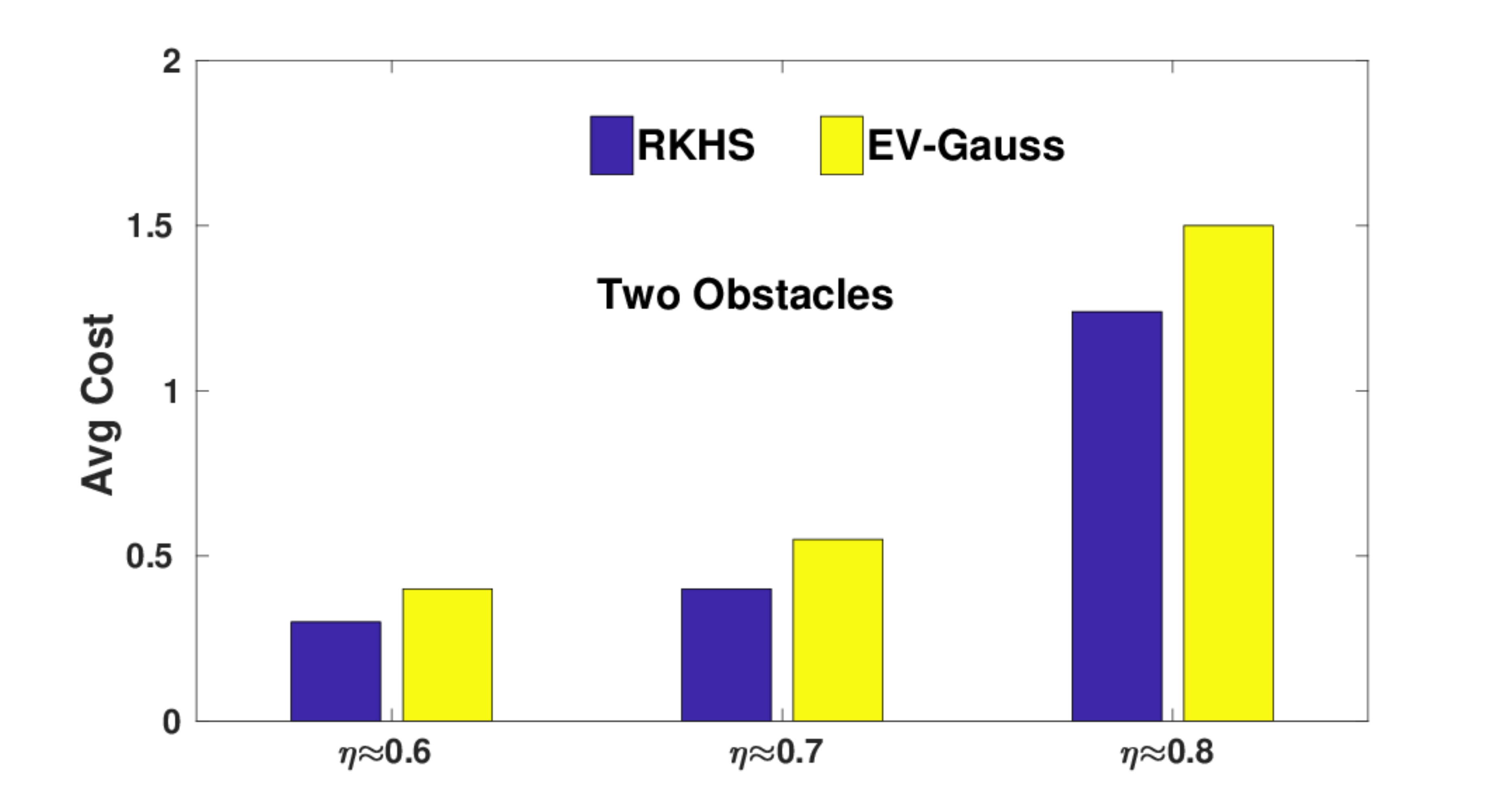}
    }
    \subfigure[]{
        \centering
        \includegraphics[width=0.3\textwidth]{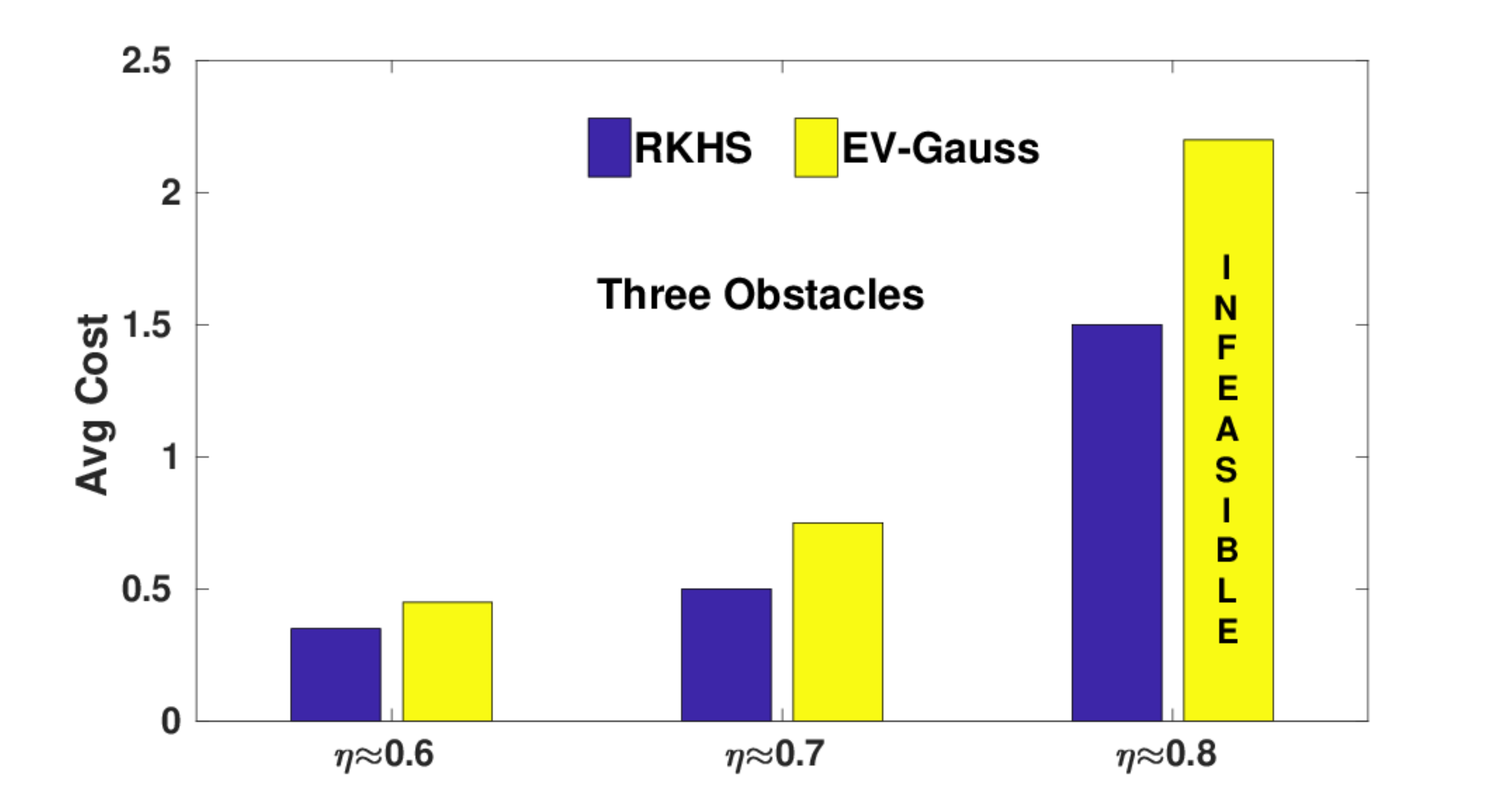}
    }
    \caption{Average cost obtained with our RKHS based algorithm and EV-Gauss. At $\eta=0.80$, EV-Gauss often resulted in infeasible solutions; The last figure (from the left) highlights this fact.}
    \label{fig:rebuttal-vs-evgauss-costs}
\end{figure*}

An example simulation is presented in the Fig. \ref{fig:rebuttal-vs-evgauss} where a robot tries to avoid three dynamic obstacles. Two different instances of this simulation were performed, that corresponded to robot requiring to  avoid the obstacles with atleast $\eta=0.60$ and $\eta = 0.80$. Fig. \ref{fig:rebuttal-vs-evgauss-confidence}, perform the validation by showing the plot of actual $\eta$ over time obtained through Monte-Carlo sampling. As can be seen, at low $\eta$, both EV-Gauss and our RKHS based algorithm result in similar $\eta$ profile. However, at high $\eta$(Fig.\ref{fig:rebuttal-vs-evgauss-confidence}(b)), EV-Gauss performs unreliably and does not maintain the required $\eta$ for the entire trajectory run. This can be attributed to the Gaussian approximation of the uncertainty used by EV-Gauss. Figure \ref{fig:rebuttal-vs-evgauss-costs} presents the comparison of average optimal cost obtained with EV-Gauss and our RKHS based approach over 10 simulation instances with varied number of obstacles. Note that, the optimal cost is a combination of norm of the tracking error and magnitude of the control inputs used. To put it quantitatively, EV-Gauss resulted in $27\%$ larger cost than our  RKHS approach $\eta=0.60$. This number jumps to $34\%$ and $40\%$ at $\eta=0.70$ and $\eta=0.80$ respectively. Furthermore, for
the three obstacle scenario at $\eta=0.80$, we were unable to
consistently obtain feasible solutions with EV-Gauss.

\bibliographystyle{IEEEtran}  
\bibliography{IEEEabrv,references} 

\end{document}